\documentclass[aps,prd,twocolumn,superscriptaddress,showpacs,amsmath,amssymb,amsfonts,amsthm,nofootinbib]{revtex4-2}
\usepackage{enumitem} 
\usepackage{mathrsfs}
\usepackage[font=small]{caption}
\usepackage{subcaption} 
\usepackage[usenames, dvipsnames]{color} 
\usepackage{graphicx}    
\usepackage{ragged2e}
\usepackage[normalem]{ulem}
\usepackage[utf8]{inputenc}
\usepackage{url}
 \usepackage{xcolor}

\usepackage{hyperref}

\begin{document}

\title{Heat engines for scale invariant systems dual to black holes}

\author{Nikesh Lilani}


\affiliation{National Institute of Technology, Rourkela, India}

\author{Manus R. Visser}

\email{manus.visser@ru.nl}

\affiliation{Institute for Mathematics, Astrophysics and Particle Physics, 
and Radboud Center for Natural Philosophy, \\Radboud University,
6525 AJ Nijmegen, The Netherlands}

\begin{abstract}
According to holography, a black hole is dual to a thermal state in a strongly coupled quantum system. One of the best-known examples of holography is the Anti-de Sitter/Conformal Field Theory (AdS/CFT) correspondence. Despite extensive work on holographic thermodynamics, heat engines for CFT thermal states have not been explored. We construct reversible heat engines where the working substance consists of a static thermal equilibrium state of a CFT. For thermal states dual to an asymptotically AdS black hole, this yields a realization of Johnson’s holographic heat engines. We compute the efficiency for a number of idealized heat engines, such as the Carnot, Brayton, Otto, Diesel, and Stirling cycles. The efficiency of most heat engines can be derived from the CFT equation of state, which follows from scale invariance, and we compare them to the efficiencies for an ideal gas. However, the Stirling efficiency for a generic CFT is uniquely determined in terms of its characteristic temperature and volume only in the high-temperature or large-volume regime. We derive an exact expression for the Stirling efficiency for CFT states dual to AdS-Schwarzschild black holes and compare the subleading corrections in the high-temperature regime with those in a generic CFT.
\end{abstract}

\maketitle

\noindent \textbf{Introduction.}  
Heat engines form a central topic in thermodynamics and played a pivotal role in its historical development \cite{wrangham1942theory,sandfort1979heat,Bharatha1989TheCA,sonntag2003fundamentals,senft2007mechanical}. A heat engine consists of a system (working substance) that converts heat into work and operates in a thermodynamic cycle. In such a   cycle,  an amount of heat ($Q_{\text{in}}$)  is  supplied  from a heat source to the system, part of  which is   converted into work ($W$) performed  on a work output device, and the remainder waste heat ($Q_{\text{out}}$) is expelled from the system  to a heat sink (we define these three quantities to be positive). The heat source and sink can be any external systems that supply and absorbs heat, respectively, but here we take them to consist of one   or more thermal reservoirs, which are large enough to exchange finite amounts of heat without changing their temperature. 

The operation of a heat engine is constrained by the first and second law of thermodynamics.  The first law, expressing energy conservation, reads: $ Q_{\text{in}} = W + Q_{\text{out}}$. Historically,    Carnot \cite{alma99362353502466} gave the earliest formulation of the second law in terms of engine efficiency.  The efficiency of a heat engine is defined as the ratio of the work done by the system and the heat supplied into the system: \begin{equation} \eta = \frac{W}{Q_{\text{in}}}= 1 - \frac{Q_{\text{out}}}{Q_{\text{in}}}\,,\end{equation}where the final expression follows from the first law. 

Carnot’s theorem   consists of two parts. 
First,   for two reservoirs at fixed temperatures 
 $T_{\text{h}}$ (hot) and $T_{\text{c}}$
  (cold), no engine can exceed the efficiency of a reversible engine operating between them. Here, reversible means recoverable: the cycle can be run backward, restoring the working substance and all surroundings (including both reservoirs) to their initial states without net change. Because the reservoirs maintain fixed temperatures, recoverability implies that the working substance be at the same temperature as the reservoir during any heat exchange, making these processes isothermal.  In this special case of two fixed-temperature reservoirs, recoverability coincides with the thermodynamic (textbook) definition of reversibility: quasi-static and free of entropy production. Second,  all such reversible engines --      exchanging heat only   isothermally with the  two reservoirs -- attain the same efficiency,  independent of the working substance or   cycle details. This universal value is known as the Carnot efficiency: $ \eta_{\text{Carnot}}= 1- T_{\text{c}}/T_{\text{h}}$, which, by the second law, is an upper bound for any irreversible engine operating between   the same two reservoirs: $\eta\leq\eta_{\text{Carnot}}$.


The efficiency of an idealized heat engine that is reversible in the thermodynamic sense (quasi-static at all stages and no entropy production) is determined by the cyclic path that the working substance traces in thermodynamic state space, which differs between engines and depends on the type of working substance. In textbooks, the ideal gas is typically used as an example for computing efficiencies of such idealized cycles, but other working substances -- such as a Van der Waals fluid \cite{madakavil2017heat} or a magnetic material \cite{KARLE2001834} -- are also possible. In this work we consider a working substance consisting of a static, global thermodynamic equilibrium state of a conformal field theory, i.e., a quantum field theory with conformal symmetry.

 
Our motivation for studying such heat engines comes from holography \cite{tHooft:1993dmi,Susskind:1994vu}, i.e.,
the idea that a gravitational theory in a $(D+1)$-dimensional spacetime is equivalent to a quantum gauge   theory without gravity living on the $D$-dimensional boundary of the spacetime. The best understood example of such a gauge/gravity duality is   the  AdS/CFT   correspondence~\cite{Maldacena:1997re,Gubser:1998bc,Witten:1998qj,Aharony:1999ti}.
A thermal high-energy state in a holographic CFT living on the (conformal) boundary of asymptotically AdS spacetime is dual to a black hole in the bulk geometry~\cite{Hawking:1982dh,Witten:1998zw}. 
Therefore, CFT heat engines are   a tool to probe black hole physics with a thermodynamic non-gravitational   system. 

Theorizing about black holes as heat engines \cite{Geroch,Bekenstein:1973ur,SCIAMA1976385,PhysRevD.43.340,Landsberg1992,Richterek,Curiel:2014zua,Bravetti:2015xsp,Prunkl:2019wdw} and interpreting black hole heat engines in terms of a dual holographic field theory \cite{Johnson:2014yja,Johnson:2016pfa,Chakraborty:2016ssb,Johnson:2019olt} is a common theme in the literature. Particularly, the idea of holographic heat engines has been proposed by Clifford Johnson \cite{Johnson:2014yja}.
We offer a conceptually distinct realization of this idea, which, to our knowledge, has not been explored before. There are   fundamental differences between Johnson's heat engines and those considered in   our work. On the one hand, Johnson's starting point is an extended version of the thermodynamics of black holes in the bulk where the cosmological constant $\Lambda$ is allowed to vary \cite{Kastor:2009wy,Dolan:2010ha,Dolan:2011xt,Cvetic:2010jb,Kubiznak:2014zwa} (see \cite{Kubiznak:2016qmn} for a review). That is, he employs a bulk pressure that is proportional to $\Lambda$ and inversely proportional to Newton's constant $G$, $P_{\text{bulk}} = - \Lambda / (8\pi G)$, and defines the thermodynamic volume as the conjugate thermodynamic quantity. On the other hand, we construct heat engines in the boundary theory, and define the pressure and volume in the   CFT in the standard thermodynamic way. It is important to mention that the bulk pressure is not dual to the CFT pressure. In fact, the bulk pressure corresponds to a   central charge $C$ in the CFT or the number of colors $N$ in a large-$N$ $\mathrm{SU(}N\mathrm{)}$  strongly-coupled gauge theory. It is questionable whether $C$ is a thermodynamic variable, since varying it changes the physical theory \cite{Mancilla:2024spp}. We keep the central charge fixed, so this is not an issue in the present work. Moreover, we stress that even though we define the heat engine in the boundary CFT, there is a  one-to-one correspondence between black hole thermodynamics \cite{Bekenstein:1973ur,Bardeen:1973gs,Hawking:1975vcx,Hawking:1982dh} and CFT thermodynamics~\cite{Witten:1998zw}. In this work we will use the   recently developed  holographic dictionary in \cite{Visser:2021eqk,Cong:2021jgb,Ahmed:2023snm,Ahmed:2023dnh,Gong:2023ywu} to compute the efficiency for heat engines dual to AdS black holes.

Our aim     is to present a   construction of holographic heat engines and to compute the efficiencies of various idealized engines: Carnot, Brayton, Otto, Diesel, Stirling and the rectangular pressure-volume cycle. We show for most heat engines, except for Stirling, the efficiency is uniquely determined by the  CFT  equation of state, and is hence the same for holographic and non-holographic CFTs. The Stirling efficiency is only fixed in terms of its characteristic temperature and volume in the high-temperature or large-volume regime, and we compare the subleading corrections in  this regime for a generic CFT and for a holographic CFT. \\

\noindent 
\textbf{CFT heat engines.} We consider heat engines whose working substance is a static, global thermodynamic equilibrium state of a CFT in $D$ spacetime dimensions. The working substance traces a closed thermodynamic cycle, returning to its initial state. We assume the cycle consists of processes that are  reversible in the thermodynamic sense: they proceed quasi-statically, so the system remains in equilibrium throughout, and they produce no   entropy.  

 For such quasi-static processes, the first law of thermodynamics reads
\begin{equation}
\label{quasistatic}
    \Delta E = Q - P \Delta V
    \qquad (\text{quasi-static}),
\end{equation}
where $Q$ is positive when heat enters the system and negative when it leaves. We hold fixed all other conserved quantities (such as electric charge or angular momentum) as well as the central charge of the CFT. The heat source and sink are modeled as (one or more) thermal reservoirs, large enough that their temperatures remain constant during heat exchange; this allows us to work with finite heat and work transfers ($\Delta$) rather than infinitesimals. 

 Because the processes   are reversible in this sense, Clausius' relation holds,
\begin{equation}
\label{eqn:heat_}
    Q = T \Delta S
    \qquad (\text{reversible}),
\end{equation}
so adiabatic and isentropic processes coincide. We can thus regard the internal energy as a function of entropy and volume, $E = E(S,V)$, suppressing dependence on other fixed parameters. For a CFT at finite temperature and in a finite volume, $E$ and $S$ are not extensive, i.e., $E(aS,aV) \neq a\,E(S,V)$; however, in the high-temperature or large-volume limit, extensivity is recovered (see below).

For the idealized heat engines that we study the cycle consists of  four paths and each path corresponds to a particular thermodynamic process, such as adiabatic $(Q=0)$, isochoric $(\Delta V=0)$, isobaric ($\Delta P=0$), and    isothermal ($\Delta T=0$) processes. Depending on the type  of processes that constitute the cycle, there are different types of heat engines. 
We label the vertices of the four paths by $i=1,2,3,4$, and $A_{i}$ denotes the value of the thermodynamic variable $A$ at the $i^{\text{th}}$ vertex. 

  In order to compute the efficiencies of various CFT heat engines, we will make use of the scale invariance of   CFTs. For homogeneous systems, scale invariance implies that the equation of state is $E= (D-1)PV $,   often called the conformal equation of state. Note that  an ideal gas system satisfies a similar equation  as a  CFT,  given by $E = \frac{f}{2}PV$, which holds in any number of dimensions. This equation   follows from combining the standard equation of state for an ideal gas $PV = N  T $  and the equipartition theorem $E=\frac{f}{2}NT $ (in units $k_B=1$),  where $f$ is the number of degrees of freedom of the gas. 
  For example, for a monatomic gas $f=D-1$ and  for a diatomic gas $f=2D-3$. Note that the CFT and ideal gas equations of state are the same if $f= 2 (D-1)$, which occurs, for instance, for a triatomic $(f=6)$ ideal gas in $D=4.$

  In order to compare the CFT and ideal gas   engines, we treat the two cases simultaneously  and represent their linear equations of state, collectively, as 
\begin{align}
\label{eqn:eqn_of_state}
     E &= \alpha PV\,,
\end{align} with $ \alpha = f/2$ for an ideal gas and $\alpha = D-1$ for a CFT.
Further, for adiabats the following relation holds 
\begin{equation}
\label{eqn: adiabat_relation}
    PV^{\frac{\alpha + 1}{\alpha}} = \text{const.} \qquad (\text{adiabat})\,.
\end{equation}
In the case of an ideal gas the exponent is $(\alpha + 1)/ \alpha = 1 + 2/f$, which is equal to the ratio $\gamma\equiv C_P/ C_V>1$ of the (temperature independent) heat capacities at constant pressure and constant volume.   \\

\noindent \textbf{Efficiencies of CFT heat engines.} We now summarize our results for the efficiencies of various CFT   heat engines. Appendix \ref{appa} in the Supplemental Material contains more detailed derivations. We express the efficiencies in terms of the characteristic thermodynamic variables of the engines that are kept fixed along the thermodynamic cycles. For the ideal gas our expressions for the efficiencies are consistent with the literature,   e.g., \cite{Shaw2008ComparingCS,Callen:450289,schroeder2000introduction, Boles2009ThermodynamicsA,sonntag2003fundamentals}.

\begin{figure}[t]
    \centering
    \includegraphics[width=0.65\linewidth]{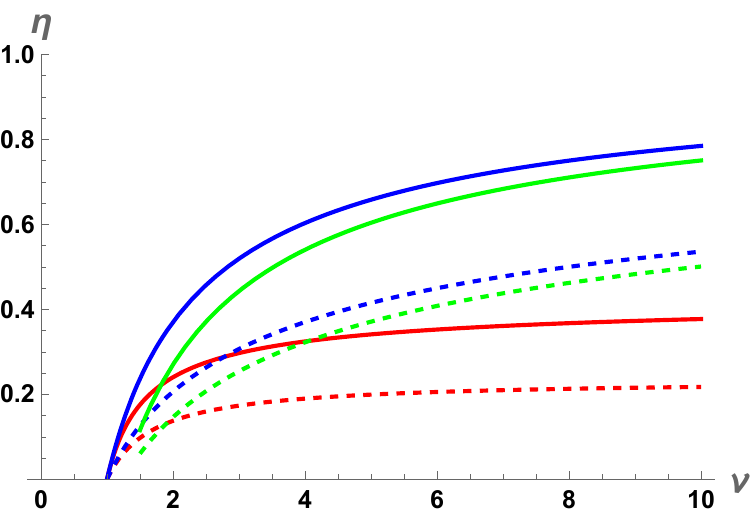}

    \caption{\emph{Efficiency   vs. compression ratio}. This plot shows the efficiency $\eta$ as a function of compression ratio $v$  for  Otto (blue),  Diesel (green),  and Stirling (red) engines in $D=4$. The solid lines correspond to a monatomic ideal gas ($f=3$)   and dashed lines   to general CFTs (for Stirling: CFT on a plane). The fixed temperature ratio for  Stirling   is $t\equiv T_{\text{h}}/T_{\text{c}} = 2$ and the fixed cutoff ratio for  Diesel   is $V_3/V_2 = 1.5$.  }
    \label{fig:figure1}
\end{figure}

A \emph{Carnot} cycle consists of isothermal expansion ($1 
\to 2$), adiabatic expansion ($2 \to 3$), isothermal compression
($3 \to 4$) and adiabatic compression ($4 \to 1$). There is an inward heat flow from the hot reservoir to the system along path $1 \rightarrow 2 $ and an outward heat flow to the cold sink along   $3 \rightarrow 4$. The  Carnot efficiency  is 
\begin{equation}
    \label{eqn: carnot_eff1}
    \eta_{{\text{Carnot}}} = 1 - \frac{T_{\text{c}}}{T_{\text{h}}}\,. 
\end{equation}
In the \emph{Brayton (or Joule)} cycle the working substance is first  compressed adiabatically ($1\to 2$), heated up isobarically $(2 \to 3)$, expanded adiabatically $(3 \to 4)$ and cooled isobarically $(4 \to 1)$.   
The Brayton efficiency is\begin{equation}
    \label{eqn: brayton_eff1}
     \eta_{\text{Brayton}} =  1- \left(\frac{P_{1}}{P_{2}}\right)^\frac{1}{1+\alpha}\,.
\end{equation}
In an \emph{Otto} cycle, which is a rough approximation of a gasoline engine, the working substance is first compressed adiabatically ($1\rightarrow 2$), then heated up isochorically ($2\rightarrow 3$),   expanded adiabatically ($3\rightarrow 4$), and finally cooled isochorically
($4\rightarrow 1$). The Otto efficiency is
\begin{equation}
    \label{eqn: otto_eff1}
    \eta_{\text{Otto}}  = 1 - \left ( \frac{V_{2}}{V_{1}} \right )^{\frac{1}{\alpha}}\,.
\end{equation}
The \emph{Diesel} cycle consists of adiabatic compression $(1 \to 2)$, isobaric heating up $(2 \to 3)$, adiabatic expansion $(3 \to 4)$ and then isochoric cooling ($4 \to 1$). 
In terms of the compression ratio $V_1/V_2$ and cutoff ratio $V_3/V_2$ the Diesel efficiency reads
\begin{equation}
    \label{eqn: diesel_eff1}
    \eta_{\text{Diesel}} =1 - \frac{\alpha}{\alpha + 1}\left ( \frac{V_{2}}{V_{1}} \right )^{\frac{1}{\alpha}}\frac{\left ( \frac{V_{3}}{V_{2}} \right )^{\frac{\alpha +1}{\alpha}}-1}{\left ( \frac{V_{3}}{V_{2}} \right )-1}\,.
\end{equation}
The efficiency of the Diesel cycle is always less than that of the Otto cycle if $V_3>V_2$, for a given compression ratio (see also Figure \ref{fig:figure1}).

For the cycle that forms a \emph{rectangle} in a $PV$-diagram, paths $2\rightarrow 3$ and $4\rightarrow 1$ are isobars, and paths $1\rightarrow 2$ and $3\rightarrow 4$  are isochores. 
The efficiency for this cycle is
\begin{equation}
\label{eqn:rectangular}
    \eta_{\text{rectangular}} = \frac{1}{(\alpha + 1)\left ( \frac{P_{2}}{P_{2}-P_{1}} \right ) + \alpha\left ( \frac{V_{1}}{V_{4}-V_{1}} \right )}\,.
\end{equation}
Note that for $  \gamma >\frac{D}{D-1} $  the Brayton, Otto and rectangular engines are more efficient for ideal gases than for CFT working substances (see  Figure \ref{fig:figure1} for the Otto engine).  

The \emph{Stirling} cycle   consists of two isothermal paths (expansion along 
$1 \to 2$ and compression along $3\to 4$),  and two isochores ($2 \to 3$ and $4 \to 1$). 
In the absence of a  regenerative heat exchanger there is  heat gain along paths $1 \rightarrow 2$ and $ 4 \rightarrow 1$, and  heat rejection along the paths $2 \rightarrow 3 $ and $3 \rightarrow 4 $. 
Without regeneration the Stirling efficiency for an ideal gas and generic CFT is
\begin{equation}
    \label{eqn: stirling_eff1a}
    \eta_{\text{Stirling}} = 1 - \frac{T_{\text{c}}(S_{3}-S_{4})+ \alpha V_{2}(P_{2}-P_{3})}{T_{\text{h}}(S_{2}-S_{1})+ \alpha V_{1}(P_{1}-P_{4})}\,.
\end{equation}
This is a universal expression that holds for a generic scale invariant system, however it depends on four thermodynamic variables, in contrast to the efficiencies of other engines. This is because in the non-regenerative Stirling cycle  there is heat exchange along all four  paths, and the heat exchanges along the isotherms and isochores  cannot be expressed in terms of the same thermodynamic variables. We want  to express the efficiency \eqref{eqn: stirling_eff1a} in  terms of   $T$ and $V$ alone, which are the characteristic parameters  of the Stirling engine, since they are constant along the isotherms and isochores, respectively, and they are experimentally controllable. In order to so do we need to know the functions $S(T,V)$ and $P(T,V)$, which  depend on   the details of the CFT and the spatial geometry. 

For concreteness, we now consider a CFT working substance with a characteristic scale~$R$ and volume $V\propto R^{D-1}$, such as   a round sphere of radius $R$.  
The dimensionless products $ER$ and $TR$ are then scale invariant, which    do not change as one varies the volume. This implies the entropy and dimensionless energy $ER$ only depend on $T$ and $V$ via the product $TR.$ In a high-temperature or large-volume expansion of the entropy and energy the leading term is extensive, i.e., $S \propto (TR)^{D-1}\propto T^{D-1} V$ and $ER \propto (TR)^D$, or, equivalently, $E \propto T^D V$ \cite{Verlinde:2000wg}. The pressure   follows from inserting the scaling of the energy into the conformal equation of state, yielding $P\propto T^D$. Moreover, the scaling of the subleading terms  in an expansion around $TR=\infty$ is also fixed: the next order is always subleading in $(TR)^{-2}$ with respect to the previous order. For instance, the expansion  of  the scale invariant product of the canonical free energy $F$ and $R$ in    any  CFT is \cite{Kutasov:2000td}
\begin{equation}
\label{freeenergy}
- FR = a_D (2\pi TR)^D + a_{D-2} (2\pi TR)^{D-2} + \mathcal O((TR)^{D-4})\,.
\end{equation} 
From this expansion  the entropy and pressure can be explicitly computed via the standard thermodynamic relations $S=-(\partial F/ \partial T)_V$ and $P=-(\partial F/ \partial V)_T$, see Appendix \ref{appc} in the Supplemental Material. Inserting this into the Stirling efficiency \eqref{eqn: stirling_eff1a} yields 
\begin{equation}
    \label{eqn:eff_CFT_on_plane_uncharged}
    \eta_{\text{Stirling}}^{\text{CFT}} = 1 - \frac{T_{\text{c}}^{D}(V_{2}\xi_{12}-V_{1}\xi_{11})+V_{2}\frac{D-1}{D}(T_{\text{h}}^{D}\chi_{22}-T_{\text{c}}^{D}\chi_{12})}{T_{\text{h}}^{D}(V_{2}\xi_{22}-V_{1}\xi_{21})+V_{1}\frac{D-1}{D}(T_{\text{h}}^{D}\chi_{21}-T_{\text{c}}^{D}\chi_{11})},
\end{equation}
where $\xi_{ij}$ and $\chi_{ij}$ are up to order $\mathcal O (T_i^{-4} V_j^{-4/(D-1)})$
\begin{align}
    \xi_{ij} &\approx 1 + \frac{a_{D-2}(D-2)}{a_D D (2\pi)^2 T_i^2} \left ( \frac{\Omega_{D-1}}{V_j} \right)^{\frac{2}{D-1}}\,, \label{xi}\\
    \chi_{ij} &\approx 1 + \frac{a_{D-2}(D-3)}{a_D (D-1)(2\pi)^2T_i^2}\left ( \frac{\Omega_{D-1}}{V_j} \right)^{\frac{2}{D-1}}\,. \label{chi}
\end{align}
Here $T_{1} \equiv T_{\text{c}}$ and $T_{2} \equiv T_{\text{h}}$. Note   the Stirling efficiency is uniquely fixed to leading order in the high-temperature or large-volume expansion. But to subleading order $\eta_{\text{Stirling}}^{\text{CFT}}$ depends on $a_{D}$ and $a_{D-2}$, which are defined via \eqref{freeenergy} as the coefficients of the leading and subleading terms in the free energy expansion. These coefficients are  independent of $(T,V)$, but do   depend on the matter content of    CFTs.   They have been explicitly computed for free CFTs in $D=4$ and $D=6$ in \cite{Kutasov:2000td}. For instance, for $\mathcal N=4$ SYM theory with $\mathrm{SU(}N\mathrm{)}$  gauge group in $D=4$ we have $a_4 = (N^2-1)/48$ and $a_2  = - (N^2-1)/8$, so $a_4 / a_2 =-1/6$.

\begin{figure}[t]
    \centering
    \includegraphics[width=0.65\linewidth]{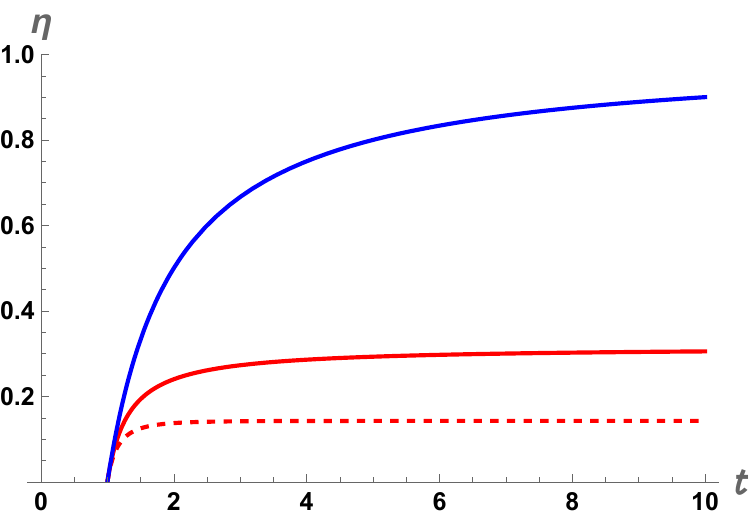}
    \caption{\emph{Efficiency   vs. temperature ratio.} This plots shows the efficiency $\eta$ as a function of  $t\equiv T_{\text{h}}/T_{\text{c}}$  for Stirling (red) and Carnot (blue) engines in $D=4$. The solid lines correspond to a monatomic  ideal gas ($f=3$) and the dashed line    to a CFT on a plane.  The fixed compression ratio for Stirling   is $v\equiv V_2/V_1= 2$. }
    \label{fig:figure2}
\end{figure}

Further, for an ideal gas the change in the entropy along an isotherm is given as $\Delta S = N \ln\left(V_2/V_1\right)$ and the pressure is related to the temperature and volume by the equation of state $P  = NT /V $.   Hence, the Stirling efficiency for an ideal gas is given by
\begin{equation}
    \label{eqn: stirling_IG}
    \eta_{\text{Stirling}}^{\text{ideal gas}} = 1 - \frac{T_{\text{c}}\ln(V_2/V_1) + \frac{f}{2}(T_{\text{h}}-T_{\text{c}})}{T_{\text{h}}\ln(V_2/V_1) + \frac{f}{2}(T_{\text{h}}-T_{\text{c}})}\,.
\end{equation}
We thus find that the dependence of the Stirling efficiency on $T$ and $V$ is different for an ideal gas and a CFT working substance.

Although all the heat cycles we consider are reversible in the thermodynamic sense, their efficiencies are not constrained by the second part of Carnot’s theorem (see Introduction) to equal the Carnot efficiency -- except for the Carnot cycle itself. This part of the theorem applies only to engines that are recoverable, which for two fixed-temperature reservoirs means all heat exchange must be isothermal with the appropriate reservoir. The Brayton, Otto, Diesel, and rectangular cycles fail this condition, as they contain no isothermal steps; heat is transferred along non-isothermal paths where the working substance's temperature changes. Even if such cycles are carried out quasi-statically and without entropy production, these steps would require a continuum of reservoirs to remain reversible, violating the two-reservoir assumption. The Stirling engine does include isothermal expansion and compression with fixed-temperature reservoirs, but also has isochores where the working substance's temperature changes, making those steps non-isothermal and likewise non-recoverable. The Carnot cycle alone consists entirely of isotherms and adiabats, is recoverable, and satisfies all the assumptions of the theorem, thus achieving $\eta_{\text{Carnot}} = 1 - T_{\text{c}}/T_{\text{h}}$.

In Figure \ref{fig:figure1} we plotted  the  efficiencies as a function of   compression ratio $v$ (the ratio of larger volume and smaller volume, i.e., $V_1/V_2$ for Otto and Diesel and $V_2/V_1$ for Stirling) for the Otto, Diesel and Stirling engines. For Stirling the temperature ratio $t\equiv T_{\text{h}}/T_{\text{c}}$ is kept fixed and for Diesel the cutoff ratio $V_3/V_2$ is fixed. The Otto and Diesel efficiencies asymptote to 1, and the $v\to \infty$ limit of the Stirling efficiency    is  $1-t^{-1}$ for an ideal gas and $(1 - t^{-D})/D$ for a CFT on a plane. In Figure~\ref{fig:figure2} we plotted the efficiencies as a function of the temperature ratio $t$, at fixed compression ratio $v$, for the Carnot and Stirling engines. These plots show that  the efficiency is universally higher for (monatomic) ideal gases than for CFTs.    The Carnot efficiency asymptotes to 1, and the Stirling efficiency asymptotes to $(v-1)/(Dv-1)$ for a CFT on a plane and  to $ \ln(v)/(f/2+  \ln(v))$ for an ideal gas.

\begin{figure*}[ht]
    \centering
    \begin{subfigure}[b]{0.28\textwidth}
        \centering
        \includegraphics[width=\textwidth]{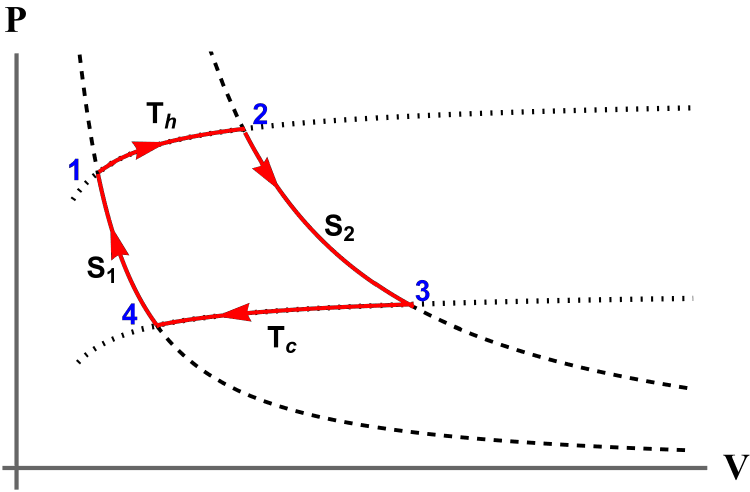} 
        \caption*{(a) Carnot engine }
        \label{fig:1a}
    \end{subfigure} \hspace{10mm}
    \begin{subfigure}[b]{0.28\textwidth}
        \centering
        \includegraphics[width=\textwidth]{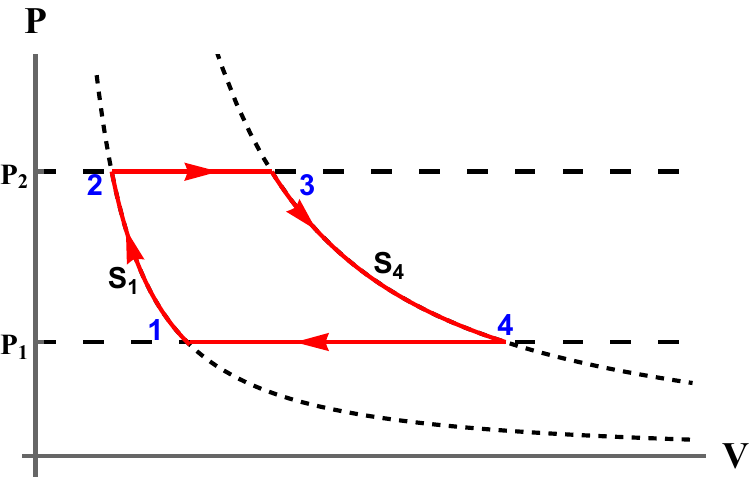} 
        \caption*{(b) Brayton engine }
        \label{fig:1b}
    \end{subfigure}  \hspace{10mm}
    \begin{subfigure}[b]{0.28\textwidth}
        \centering
        \includegraphics[width=\textwidth]{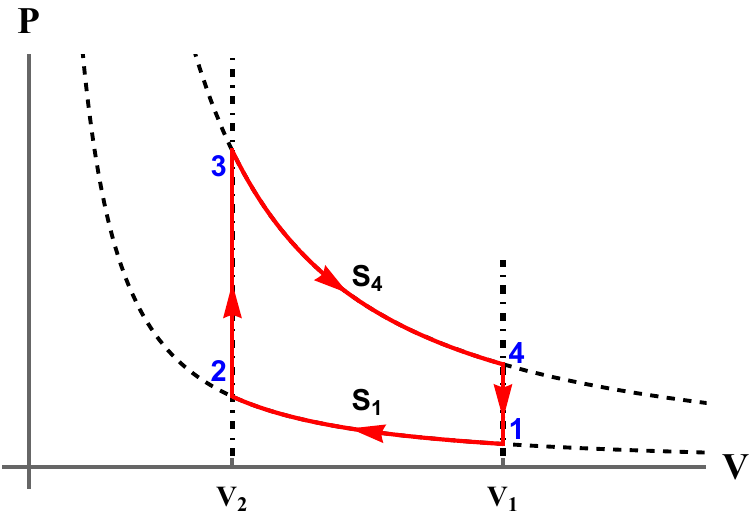} 
        \caption*{(c) Otto engine }
        \label{fig:1c}
    \end{subfigure}

    \par\bigskip 
    
    \begin{subfigure}[b]{0.28\textwidth}
        \centering
        \includegraphics[width=\textwidth]{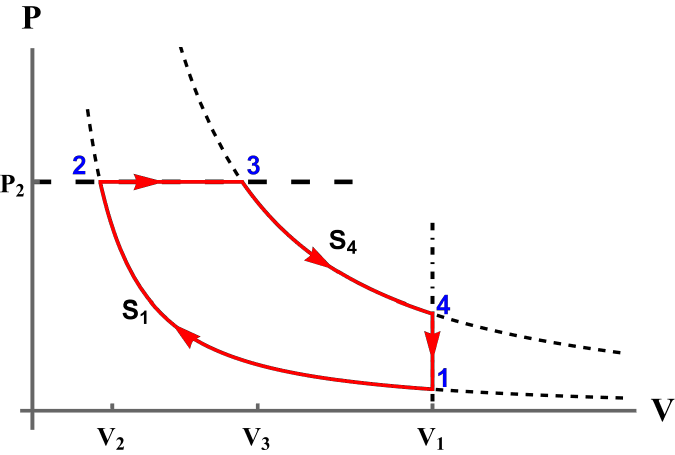} 
        \caption*{(d) Diesel engine }
        \label{fig:1d}
    \end{subfigure}\hspace{10mm}
    \begin{subfigure}[b]{0.28\textwidth}
        \centering
        \includegraphics[width=\textwidth]{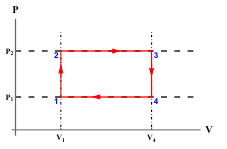} 
        \caption*{(e) Rectangle engine }
        \label{fig:1e}
    \end{subfigure}\hspace{10mm}
    \begin{subfigure}[b]{0.28\textwidth}
        \centering
        \includegraphics[width=\textwidth]{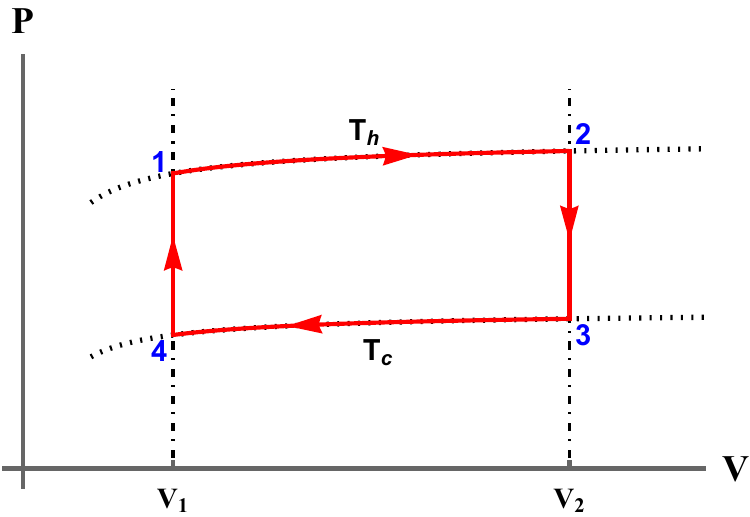} 
        \caption*{(f) Stirling engine}
        \label{fig:1f}
    \end{subfigure}

    \caption{\emph{Pressure-volume diagrams for holographic CFT heat cycles.} These plots are heat cycles for thermal CFT working substances dual to AdS-Schwarzschild black holes. The number of CFT spacetime dimensions is $D=4$. In all figures   dotted lines correspond to isotherms, short dashed lines to adiabats, long dashed lines to isobars, and dotdashed lines to isochores. The red curves indicate the cycle, the numbers at the vertices denote the ordering of the cycle,  and the arrows the direction of the cycle.   $T_{\text{h}}$ and $T_{\text{c}}$ are the temperatures of the hot and cold reservoirs, respectively. The panels represent (a) the Carnot cycle   (isotherm-adiabat-isotherm-adiabat), (b) Brayton cycle (adiabat-isobar-adiabat-isobar), (c) Otto cycle (adiabat-isochore-adiabat-isochore), (d) Diesel cycle (adiabat-isobar-adiabat-isochore),  (e)  rectangle cycle (isochore-isobar-isochore-isobar), and (f)  Stirling cycle (isotherm-isochore-isotherm-isochore).}
    \label{fig:five_figures}
\end{figure*}

\begin{figure*}[ht] 
    \centering
    \begin{subfigure}[b]{0.55\columnwidth}
        \centering
        \includegraphics[width=1.15\textwidth]{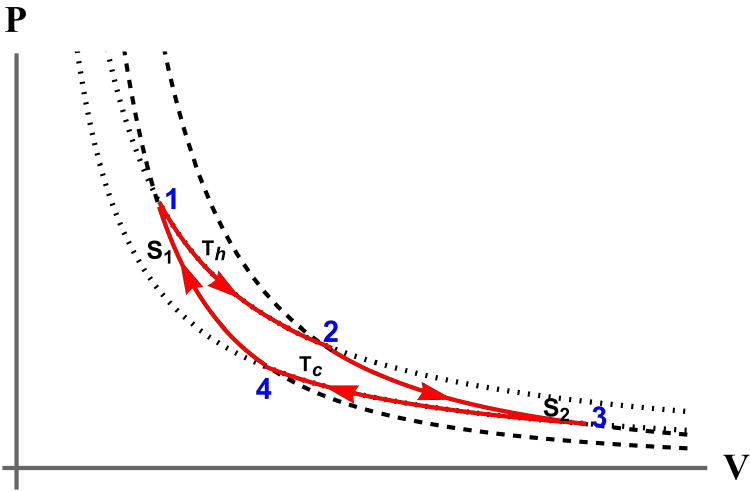} 
        \subcaption{Carnot engine} 
        \label{fig:2a}
    \end{subfigure}\hspace{10mm}
    \begin{subfigure}[b]{0.55\columnwidth}
        \centering
        \includegraphics[width=1.15\textwidth]{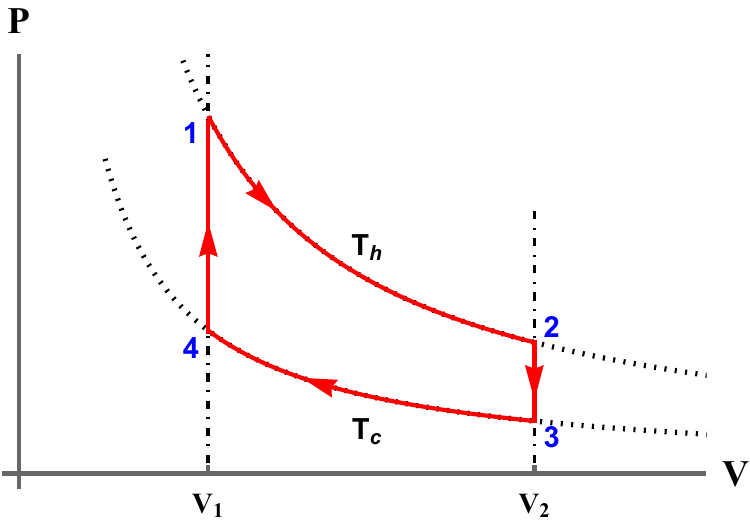} 
        \subcaption{Stirling engine} 
        \label{fig:2b}
    \end{subfigure}

    \caption{\emph{Pressure-volume diagrams for ideal gas heat cycles}. These plots are heat cycles for (a) Carnot (isotherm-adiabat-isotherm-adiabat) and (b) Stirling  (isotherm-isochore-isotherm-isochore) engines with a working substance consisting of a monatomic ($\gamma=5/3$) ideal gas in $D=4$ spacetime dimensions. The  dotted lines correspond to isotherms, short dashed lines to adiabats,  and dotdashed lines to isochores. The red curves indicate the cycle, the numbers at the vertices denote the ordering of the cycle,  and the arrows the direction of the cycle.   $T_{\text{h}}$ and $T_{\text{c}}$ are the temperatures of the hot and cold reservoirs, respectively.  }
    \label{fig:two_figures}
\end{figure*}

Finally, we plotted the $PV$-diagrams  for all  heat engines   in Figure~\ref{fig:five_figures} (for a holographic CFT) and Figure~\ref{fig:two_figures} (for a monatomic ideal gas), and the $TS$-diagrams in Figure~\ref{fig:lefT_column} (for a holographic CFT) and Figure~\ref{fig:righT_column} (for a monatomic ideal gas). In Appendix \ref{appb} of the Supplemental Material   we derive the  equations for the various  cycle paths that are used to make these plots.  The $PV$-diagrams for the CFT and ideal gas systems are identical for the Brayton, Otto,  Diesel and rectangle engines, but different for the Carnot and Stirling engine. Moreover, the $TS$-plots corresponding to the CFT and ideal gase systems are different for the Brayton, Otto, Diesel, rectangle and Stirling engines, but identical for the Carnot engine. By comparing the Carnot (Figs.~3a and 4a) and Stirling cycles (Figs.~3e and 4b)  we see that the     isotherm for a CFT monotonically increases with $V$ whereas the isotherm for the  ideal gas monotonically decreases with~$V$. The slope of the adiabats is also different for the two systems.  Further, by comparing the cycles in Figs.~\ref{fig:lefT_column} and~\ref{fig:righT_column} we see that the  isochores and isobars in a $TS$-diagram     are different for   CFT and ideal gas systems.\\

\noindent \textbf{Holographic heat engines.}
So far we have considered heat engines for generic CFTs.  Next, we construct heat engines for holographic CFT states that are dual to AdS black holes.  We stress that the generic CFT results for the engine efficiencies above also hold for holographic CFTs, but for the Stirling engine we can compute the efficiency exactly by invoking holography.  For   heat engines of holographic CFTs   the   geometry is fixed to be equivalent (up to   Weyl rescaling) to  the boundary geometry of the dual black hole spacetime. That is because  we take the working substance of holographic heat engines to be  the entire spatial geometry of the holographic  CFT. Furthermore, we only consider   black holes with positive heat capacity, since if the  heat capacity were negative the   cycles in the $PV$-diagrams \ref{fig:five_figures} and \ref{fig:two_figures} would act as refrigerators (and the reverse cycles would be heat engines). Large enough AdS  black holes  indeed have positive heat capacity and thus their   thermodynamic cycles (in the order $1 \to 2 \to 3 \to 4 \to 1$) can operate as heat engines. 

Concretely, here we consider   static, spherically symmetric, uncharged asymptotically AdS black holes, a.k.a. AdS-Schwarzschild black holes, in $D+1$ spacetime dimensions. Hence, in our setup the spatial geometry of the holographic heat engine is  a round sphere with radius~$R$ and   volume   $V=\Omega_{D-1}R^{D-1}.$   For these black holes the holographic dictionary 
reads    (see Appendix \ref{appd} in the Supplemental Material  for a derivation) 
\cite{Visser:2021eqk,Cong:2021jgb,Ahmed:2023snm,Ahmed:2023dnh}
\begin{align}
    \label{eqn:central_charge}
   S &= 4 \pi C x^{D-1}, \qquad   C = \frac{\Omega_{D-1} L^{D-1}}{16 \pi G}\,,\\
   \label{eqn:holo_cft_energy1}
    E &= \frac{(D-1)Cx^{D-2}}{R}\left (1 + x^{2} \right)\,,\\
    \label{eqn:holo_temp1}
    T &= \frac{D-2}{4 \pi R x}\left (1 + \frac{D}{D-2}x^{2} \right)\,,\\
    P&=\frac{Cx^{D-2}}{\Omega_{D-1}R^D}\left (1 + x^{2} \right)\,.\label{holo:pressure}
\end{align}
We defined    $x \equiv r_{h}/L$ with $r_{h}$   the horizon radius of the black hole  and $L$   the AdS curvature radius. The heat capacity at fixed $V$ and $C$   is positive if   $x> \sqrt{(D-2)/D}$. Crucially, the boundary volume $V$ and the central charge $C$ can be independently varied, since they depend on $R$ and $L$, respectively. In previous dictionaries, e.g., in \cite{Dolan:2014cja,Karch:2015rpa,Visser:2021eqk}, $R$ was set equal to $L$, so that  $V$ and $C$ could be independently varied only if Newton's constant $G$ is allowed to change. For holographic heat engines, however, we want to keep the theory parameters in the bulk and boundary fixed ($G$ and $C$) while allowing $V$ to vary, which is possible only if $R \neq L$ \cite{Ahmed:2023snm}.

We now compute the Stirling efficiency  by invoking the holographic dictionary above.  From  \eqref{eqn:central_charge}-\eqref{holo:pressure}   one can derive exact expressions for  $S(T,V)$ and $P(T,V)$ (see Appendix \ref{appe} of the Supplemental Material). Inserting them into   \eqref{eqn: stirling_eff1a} yields  
  that   the Stirling efficiency for CFT states dual to AdS-Schwarzschild takes the same form as \eqref{eqn:eff_CFT_on_plane_uncharged}, but now the functions $\xi_{ij}$ and $\chi_{ij}$ are given by
\begin{align}
    \label{exactxi}
    &\xi_{ij} = \frac{1}{2^{D-1}}\!\left[ 1 + \sqrt{1 - \frac{D(D-2)}{4 \pi^{2} T_{i}^{2}} \left(\frac{\Omega_{D-1}}{V_j}\right)^{\frac{2}{D-1}}}  \right]^{D-1} \,\!\!\!\!\!\!\!\!\!\!, \nonumber\\
    &\chi_{ij} = \xi_{ij} ^{\frac{D-2}{D-1}} \left[ \xi_{ij}^{\frac{2}{D-1}} + \frac{D^{2}}{16 \pi^{2} T_i^{2}} \left(\frac{\Omega_{D-1}}{V_j}\right)^{\frac{2}{D-1}} \right]\,.
\end{align}
These are exact expressions in the temperature and volume. We can expand them at high temperature or large volume. The result 
  up to subleading order   is the same as \eqref{xi} and \eqref{chi} with the ratio of the coefficients given by  \begin{equation}\label{holographiccoeff}
    \frac{a_{D-2}}{a_D} = - \frac{D^2 (D-1)}{4}\,.
\end{equation}
This agrees with earlier findings for these coefficients in holographic CFTs \cite{Verlinde:2000wg,Kutasov:2000td}. Importantly, the holographic   Stirling efficiency  is lower than   the  leading order contribution to the efficiency in the high-temperature and large-volume expansion, for which $\xi_{ij}=\chi_{ij}=1$.    Moreover, we checked by plotting that for $\mathcal N=4$ SYM theory in $D=4$ the Stirling efficiency is higher  at zero 't Hooft coupling (for which $a_4/a_2=-1/6$) than at  infinite coupling (with $a_4 /a_2=-1/12 $, cf. \eqref{holographiccoeff}). Thus, this suggests for CFTs the Stirling efficiency decreases as the coupling increases.  \\

\noindent  
\textbf{Comparison with Johnson's holographic heat engines.}  Next we compare our holographic heat engines with those in Johnson's work \cite{Johnson:2014yja,Johnson:2016pfa,Chakraborty:2016ssb,Johnson:2019olt} and subsequent follow-ups. Apart from the fact that both heat engines make use of the AdS/CFT correspondence, they are completely different. The key predictions for the efficiencies  of all heat engines  are distinct, and the way the engines operate  is also different, as we   explain below. Moreover, we do not just give   predictions for holographic heat engines, but also for generic CFTs.

The main difference between the two constructions lies in the definitions of pressure and volume. Johnson considers the bulk pressure $P_{\text{bulk}}=- \Lambda/(8\pi G)$,  proportional to the cosmological constant $\Lambda$, and defines the volume as its conjugate  quantity in the extended first law of black holes, in which $\Lambda$ is being varied \cite{Kastor:2009wy,Dolan:2010ha,Dolan:2011xt,Cvetic:2010jb,Kubiznak:2014zwa}. On the other hand, we construct heat engines in the dual thermal conformal field theory, where pressure and volume are defined in   standard thermodynamic terms. These distinct definitions of pressure and volume imply that our holographic heat engines function in an entirely different way from Johnson's engines. For instance,  Johnson \cite{Johnson:2014yja} considered charged AdS black holes  instead of Schwarzschild–AdS black holes, because in his approach the former allow for nontrivial engine cycles in the $PV$-plane, while the latter do not. Our construction already gives nontrivial cycles for Schwarzschild-AdS black holes.

A consequence of the previous point and a crucial difference is that in Johnson's heat cycles, the underlying theory changes when pressure varies, whereas in our construction, the theory remains fixed. This is because varying the bulk pressure in Johnson's approach corresponds to adjusting the cosmological constant and the number of colors $N$ in the boundary theory. Consequently, in his model, the boundary theory itself changes during the thermodynamic cycle. However, pressure should be a thermodynamic state variable, meaning it is a property of the state and not of the theory. As emphasized in \cite{Mancilla:2024spp}, $N$ is not a function of the boundary spacetime, which is required for a state variable describing local thermodynamic equilibrium. This implies that changing  $N$ does not correspond to a standard thermodynamic process, but rather a flow within the space of CFTs. Therefore, Johnson \cite{Johnson:2014yja} conjectured that his engine cycles could be realized using renormalization group flow, but it remains unclear whether this is feasible, and it stands in conflict with the usual operation of heat engines. Our holographic heat engines, on the other hand, operate in the conventional thermodynamic sense, by adjusting the thermal state quasi-statically, which is a major advantage over Johnson's approach. Furthermore, we demonstrated that the efficiency of our scale-invariant heat engines is comparable to that of ideal gas engines, showing that our approach follows standard thermodynamic principles.

A more fine-grained difference is that in Johnson's approach the Carnot and Stirling engine are identical for charged AdS black holes, since adiabats are equal to isochores, whereas in our approach they are not. This is because both the entropy and the   volume  in extended  thermodynamics  of static AdS black holes depend only on the horizon radius, so they are not independent \cite{Dolan:2012jh}. This is problematic in itself, because it implies that the   energy function $E(S,V)$ and its partial derivatives \( \left( \frac{\partial E}{\partial S}\big|_V, \frac{\partial E}{\partial V} \big|_S\right) \) are  ill defined. Our approach does not suffer from this degeneracy, since entropy and volume are independent variables. Thus, in our construction the Carnot and Stirling engine are   distinct, as they should be.

Another key advantage of our proposal is that the efficiencies of    holographic heat engines can  potentially be experimentally tested. This is possible because our working substance consists of a strongly coupled CFT thermal state, which can be realized at the quantum critical point of a condensed matter system at finite temperature, such as high-temperature superconductors (see \cite{Hartnoll:2016apf} for a review). The Stirling efficiency in \eqref{eqn:eff_CFT_on_plane_uncharged} and \eqref{exactxi} provides a distinct prediction for a thermal CFT system dual to a black hole. As a result, our work offers a  framework for experimentally testing holographic models in a condensed matter setting.
In contrast, no direct experimental connection can be made with Johnson’s heat engines, as the theory evolves along the heat cycles, and because the bulk pressure and volume do not agree with those in the CFT.  Furthermore, Johnson’s heat engines do not implement the CFT equation of state, which played a crucial role in our approach. \\

\noindent \textbf{Conclusion.} In this paper we proposed a   way to construct heat engines in holographic field theories. The working substance can be modeled by a  strongly coupled, large-$N$, conformally invariant thermal system that is   dual to a black hole spacetime \cite{Witten:1998zw,Heemskerk:2009pn}. A crucial aspect of the holographic dictionary that we used is that the volume can be independently varied from the other thermodynamic variables. We computed the efficiencies of various idealized  engines for (holographic) CFTs. 

For future work there are many   generalizations of our setup worth studying. We only described the simplest holographic heat engines as a proof of principle that our construction works. First, one could study other     types of   engines, ideally    more realistic ones for which  the efficiency of holographic   systems can be experimentally tested.  Second,  we considered only field theories with conformal symmetry, but one could define heat engines for holographic field theories with different global symmetries, such as anisotropic scaling symmetry \cite{Taylor:2015glc,Cong:2024pvs}.  Third, one could   compute the Stirling efficiency for  specific    CFTs at finite temperature and volume, for instance perturbatively at weak coupling, and compare   with the holographic result. It would be particularly valuable to investigate the coupling dependence of the Stirling efficiency and to understand the physical origin behind the higher efficiency at weak coupling.  Finally, one could study holographic   engines for different types of black holes, such as charged or rotating black holes   or black hole solutions to   higher curvature gravity. \\

\noindent \textit{Acknowledgments:} M.R.V. is grateful to S. Borsboom, E. Curiel,  T. Jacobson, K. Landsman, J. Pedraza, J. Uffink, W. Unruh,  E. Verlinde, A. Wall and D. Wallace for useful discussions. He also  thanks the participants at the Peyresq Physics 2025 conference,   where this work was presented,  for their interesting questions.  This work is supported in part by the   Spinoza Grant of the Dutch Science Organization (NWO) awarded to Klaas Landsman. \\

\newpage

\bibliography{refs}

\begin{figure*}[ht]
    \centering
    \begin{subfigure}[b]{0.55\columnwidth}
        \centering
        \includegraphics[width=1.15\textwidth]{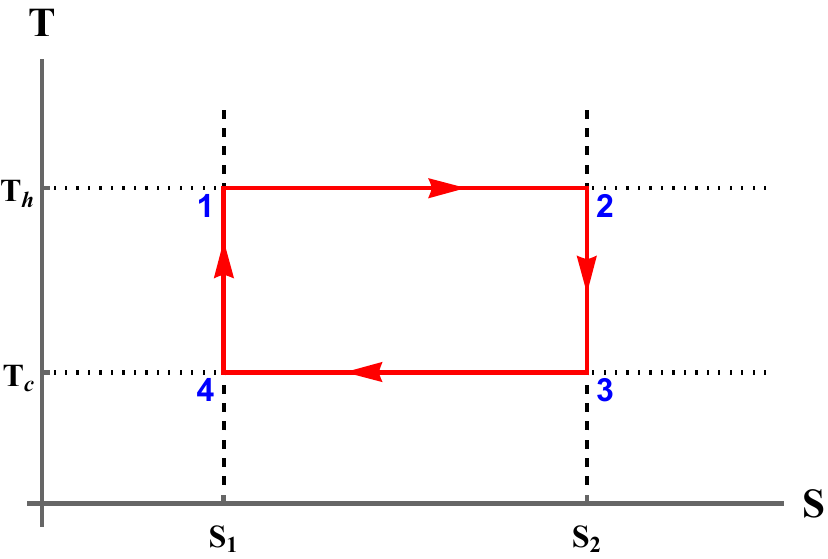} 
        \caption*{(a) Carnot engine}
    \end{subfigure}\hspace{10mm}
    \begin{subfigure}[b]{0.55\columnwidth}
        \centering
        \includegraphics[width=1\textwidth]{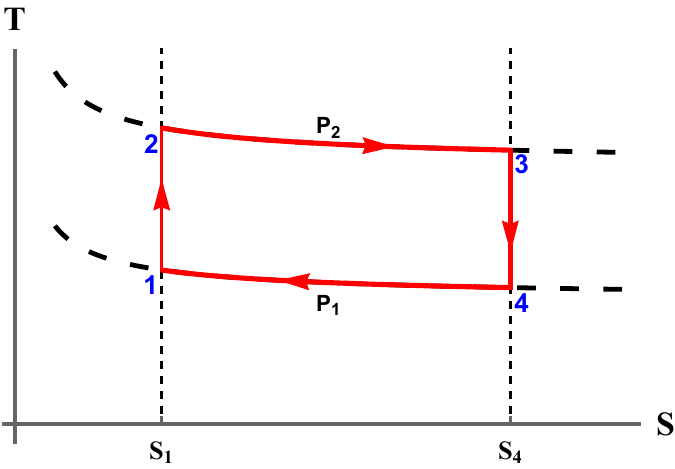} 
        \caption*{(b) Brayton engine}
    \end{subfigure}\hspace{10mm}
    \begin{subfigure}[b]{0.55\columnwidth}
        \centering
        \includegraphics[width=1.15\textwidth]{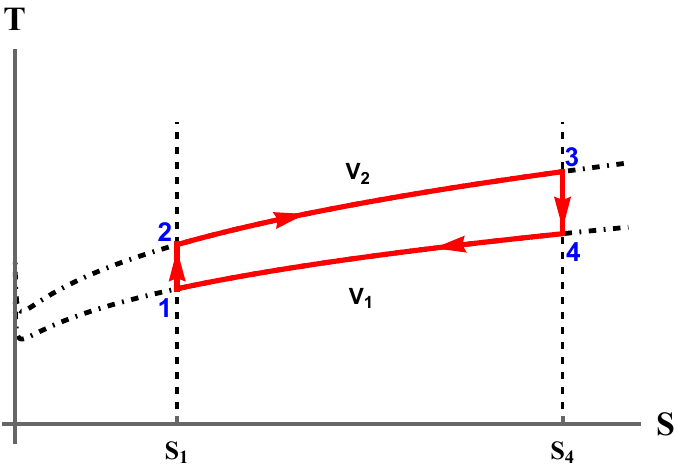} 
        \caption*{(c) Otto engine}
    \end{subfigure}
    \begin{subfigure}[b]{0.55\columnwidth}
        \centering
        
        \includegraphics[width=1.15\textwidth]{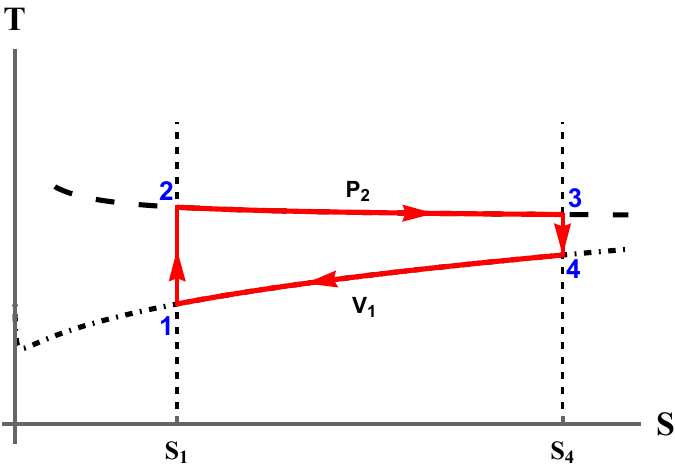} 
        \caption*{(d) Diesel engine}
    \end{subfigure}\hspace{10mm}
    \begin{subfigure}[b]{0.55\columnwidth}
        \centering
        \includegraphics[width=1.15\textwidth]{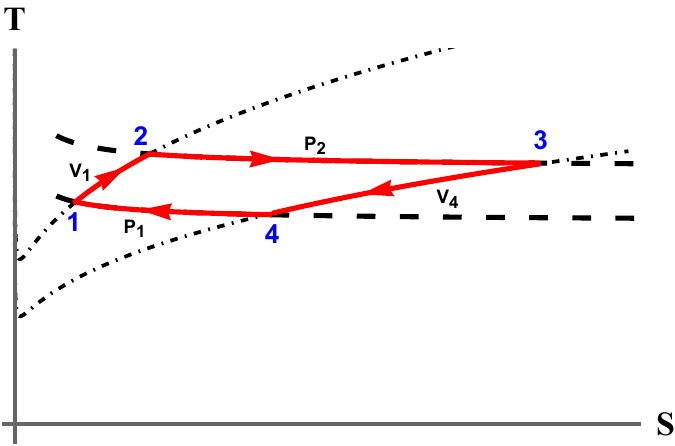} 
        \caption*{(e) Rectangle engine}
    \end{subfigure}\hspace{10mm}
    \begin{subfigure}[b]{0.55\columnwidth}
        \centering
        \includegraphics[width=1.15\textwidth]{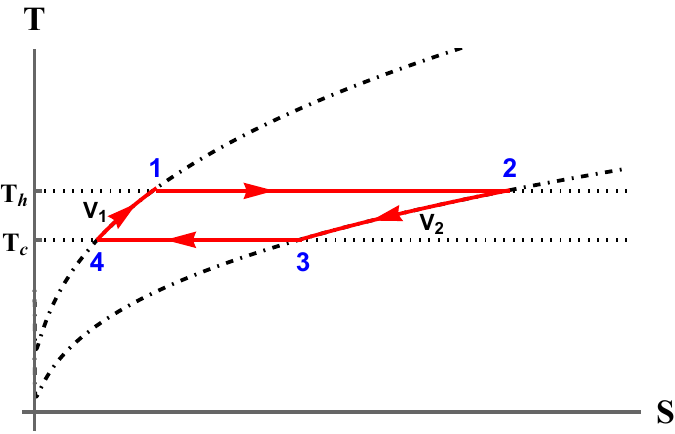} 
        \caption*{(f) Stirling engine}
    \end{subfigure}
    \caption{\emph{Temperature-entropy diagrams for holographic CFT heat cycles.} These plots are heat cycles for thermal CFT working substances dual to AdS-Schwarzschild black holes. The number of CFT spacetime dimensions is $D=4$. In all figures   dotted lines correspond to isotherms, short dashed lines to adiabats, long dashed lines to isobars, and dotdashed lines to isochores. The red curves indicate the cycle, the numbers at the vertices denote the ordering of the cycle,  and the arrows the direction of the cycle.   $T_{\text{h}}$ and $T_{\text{c}}$ are the temperatures of the hot and cold reservoirs, respectively. The panels represent (a) the Carnot cycle   (isotherm-adiabat-isotherm-adiabat), (b) Brayton cycle (adiabat-isobar-adiabat-isobar), (c) Otto cycle (adiabat-isochore-adiabat-isochore), (d) Diesel cycle (adiabat-isobar-adiabat-isochore),  (e)  rectangle cycle (isochore-isobar-isochore-isobar), and (f)  Stirling cycle (isotherm-isochore-isotherm-isochore). }
    \label{fig:lefT_column}
\end{figure*}

\begin{figure*}[ht]
    \centering
    \begin{subfigure}[b]{0.55\columnwidth}
        \centering
        \includegraphics[width=1.15\textwidth]{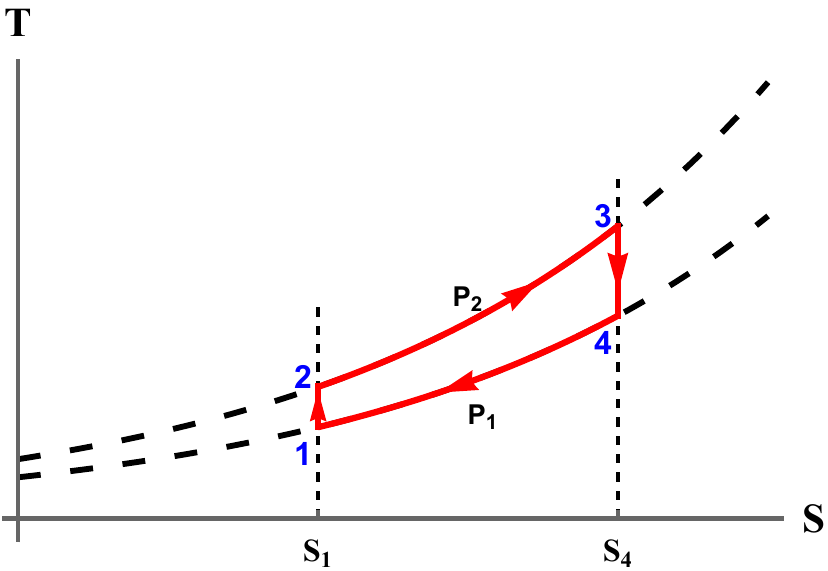} 
        \caption*{(a) Brayton engine}
    \end{subfigure}\hspace{10mm}
    \begin{subfigure}[b]{0.55\columnwidth}
        \centering
        \includegraphics[width=1.15\textwidth]{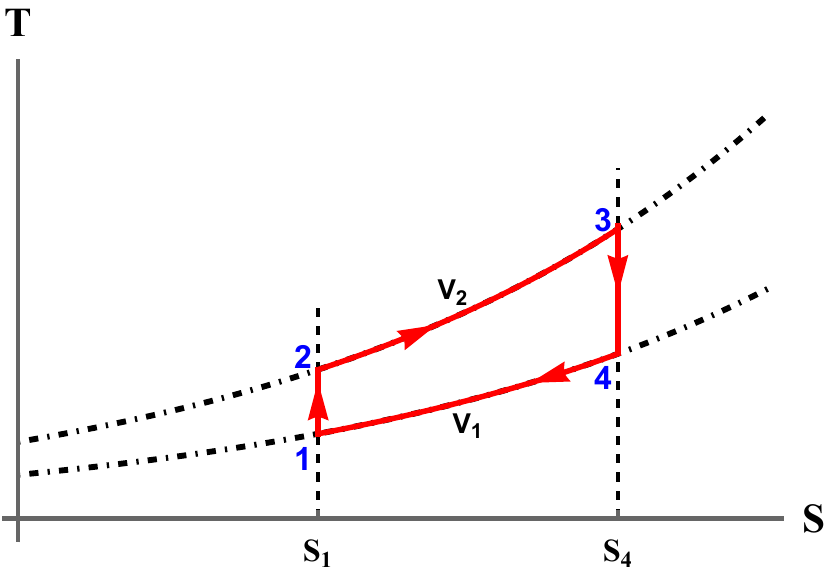} 
        \caption*{(b) Otto engine}
    \end{subfigure}\hspace{10mm}
    \begin{subfigure}[b]{0.55\columnwidth}
        \centering
        \includegraphics[width=1.15\textwidth]{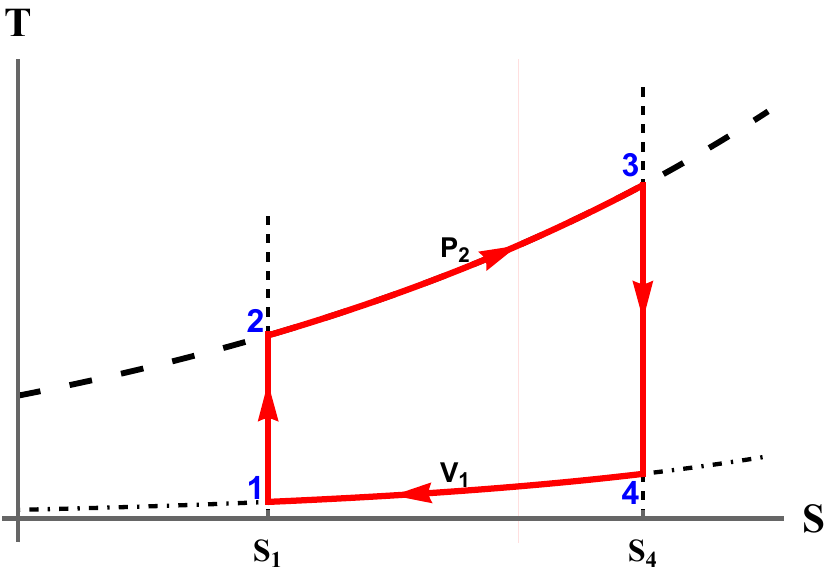} 
        \caption*{(c) Diesel engine}
    \end{subfigure} 
    \begin{subfigure}[b]{0.55\columnwidth}
        \centering
        
        \par\bigskip
        \includegraphics[width=1.15\textwidth]{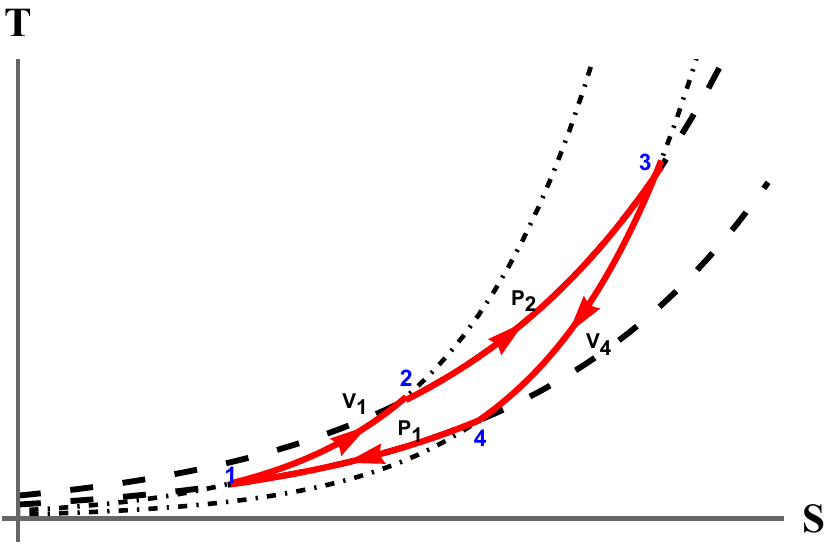} 
        \caption*{(d) Rectangle engine}
    \end{subfigure} \hspace{10mm} 
    \begin{subfigure}[b]{0.55\columnwidth}
        \centering
        \includegraphics[width=1.15\textwidth]{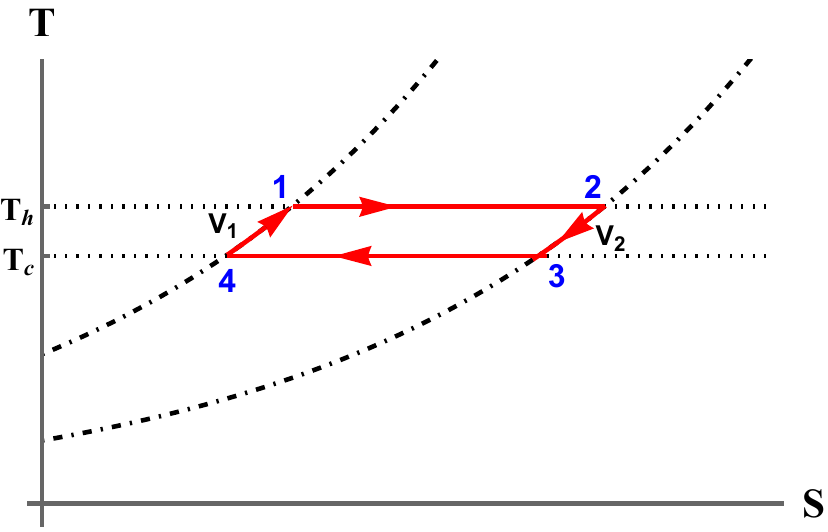} 
        \caption*{(e) Stirling engine}
    \end{subfigure}
    \caption{\emph{Temperature-entropy diagrams for ideal gas heat cycles}. These plots are heat cycles for    monatomic ($\gamma=5/3$) ideal gas working substances  in $D=4$ spacetime dimensions. The  dotted lines correspond to isotherms, short dashed lines to adiabats, long dashed lines to isobars, and dotdashed lines to isochores. The red curves indicate the cycle, the numbers at the vertices denote the ordering of the cycle,  and the arrows the direction of the cycle.  $T_{\text{h}}$ and $T_{\text{c}}$ are the temperatures of the hot and cold reservoirs, respectively.  
    The panels represent  (a) the Brayton cycle (adiabat-isobar-adiabat-isobar), (b) Otto cycle (adiabat-isochore-adiabat-isochore), (c) Diesel cycle (adiabat-isobar-adiabat-isochore),  (d)  rectangle cycle (isochore-isobar-isochore-isobar), and (e)  Stirling cycle (isotherm-isochore-isotherm-isochore). }
    \label{fig:righT_column}
\end{figure*}

\clearpage
\onecolumngrid 
\appendix

 {\begin{center}
    {\bf \Large Supplemental Material}
\end{center}

\section{Efficiencies of CFT and ideal gas heat engines}
\label{appa}
In this appendix we compute the efficiencies of various heat engines for CFT and ideal gas working substances. We recall that both thermal systems satisfy the equation of state $E=\alpha PV$, where $\alpha = D-1$ for CFTs and $\alpha = f/2=1/(\gamma-1)$ for an ideal gas with $f$ number of degrees of freedom (and $\gamma
\equiv C_P/C_V$). Regarding sign conventions,  we take the following three quantities all to be positive:   heat input $Q_{\text{in}}$ from the heat source into the system, the heat output $Q_{\text{out}}$ from the system to the heat sink and the work $W$ produced in the engine, that is done by the system on the work source. Note this is different from the sign convention in the first law, $\Delta E = Q -W$, where $Q$ is positive when it is added to the system and negative when it leaves. So with a subscript $Q_{\text{in/out}}$ is always positive and without a subscript $Q$ can be  negative. \\ 

\noindent \textbf{Carnot engine.} In a Carnot cycle the heat exchange takes place only along the two isotherms (because $ Q = 0$ along the adiabats) with an inward heat flow along $1 \rightarrow 2 $ and an outward flow along the path $3 \rightarrow 4$. From the Clausius relation \eqref{eqn:heat_} it follows that $Q_{\text{in}} = Q_{\text{in}}^{1 \rightarrow 2} = T_{\text{h}}(S_{2}-S_{1}) $ and $Q_{\text{out}} = Q_{\text{out}}^{3 \rightarrow 4} = T_{\text{c}}(S_{3}-S_{4}) = T_{\text{c}}(S_{2}-S_{1})$, since along the adiabats $2 \to 3$ and $4 \to 1$ we have    $S_{2}=S_{3}$ and $S_{1}=S_{4}$, respectively. Thus, the   efficiency of a Carnot engine   is 
\begin{equation}
    \label{eqn: carnot_eff}
    \eta_{{\text{Carnot}}} = 1 - \frac{Q_{\text{out}}^{3 \rightarrow 4}}{Q_{\text{in}}^{1 \rightarrow 2}} = 1 - \frac{T_{\text{c}}}{T_{\text{h}}}\,, 
\end{equation}
which is indeed the Carnot efficiency. Note that we did not use an equation of state   to derive this efficiency, so it holds for any working substance.\\

\noindent \textbf{Brayton engine.}
In the Brayton (or Joule) cycle there is heat exchange along the two isobaric paths, with an inward flow along the path $2 \rightarrow 3$ and an outward flow along $4 \rightarrow 1$. There is no heat exchange along the adiabatic paths   $1 \rightarrow 2 $ and $ 3 \rightarrow 4$. From the first law it follows that the heat exchange along an isobar is equal to the change in enthalpy: $Q = \Delta E + P \Delta V = \Delta (E + PV)\equiv \Delta H$. For a CFT and ideal gas the enthalpy is $ H  = (\alpha +1) PV $. Hence $Q_{\text{in}} = Q_{\text{in}}^{2\rightarrow 3} = H_{3}-H_{2} = (\alpha+1) P_{2}(V_{3}-V_{2})$ and $Q_{\text{out}} = Q_{\text{out}}^{4 \rightarrow 1} = -(H_{1}-H_{4}) = (\alpha+1) P_{1}(V_{4}-V_{1})$. The adiabat relation \eqref{eqn: adiabat_relation} for the paths $1 \to 2$ and $3 \to 4$  yields 
\begin{equation}
\label{eqn:pv_bray}
     P_{2}V_{2}^{\frac{\alpha + 1}{\alpha}} = P_{1}V_{1}^{\frac{\alpha + 1}{\alpha}}\,,\qquad
     P_{2}V_{3}^{\frac{\alpha + 1}{\alpha}} = P_{1}V_{4}^{\frac{\alpha + 1}{\alpha}} \,.
\end{equation}
Dividing these two equations implies the volume ratios are equal: $V_2/V_3 =V_1/V_4$. From this equality and  \eqref{eqn:pv_bray}  we obtain the   efficiency   for the Brayton engine
\begin{equation}
    \label{eqn: brayton_eff}
     \eta_{\text{Brayton}} = 1- \frac{ Q_{\text{out}}^{4\rightarrow 1} }{ Q_{\text{in}}^{2\rightarrow 3} } = 1 - \frac{H_{4}-H_{1}}{H_{3}-H_{2}} = 1- \left(\frac{P_{1}}{P_{2}}\right)^\frac{1}{1+\alpha}\,.
\end{equation}
The Brayton efficiency for a CFT working substance thus depends on the number of  dimensions. Note that for $  \gamma >\frac{D}{D-1} $ the ideal gas Brayton engine is more efficient than the corresponding CFT engine. For instance,  a Brayton engine consisting of a monatomic gas ($f= D-1, \gamma = (D+1)/(D-1)$) or diatomic gas ($f=2D-3,\gamma = (2D-1)/(2D-3)$)   is more efficient than  a CFT Brayton engine in $D$ spacetime dimensions.\\

\noindent \textbf{Otto engine.}
In an Otto cycle the processes along $1\rightarrow 2$ and $3\rightarrow 4$ are adiabatic and the processes along $2\rightarrow 3$ and $4\rightarrow 1$ are isochoric (which implies $V_{1} = V_{4}$ and $V_{2} = V_{3})$. There is heat loss along $4\rightarrow 1$ and heat gain along $2\rightarrow 3$. From the first law it follows that heat exchange along isochores is equal to the change in internal energy. Thus, $Q_{\text{in}} = Q_\text{in}^{2 \rightarrow 3} = E_{3} - E_{2} = \alpha V_{2} (P_{3} - P_{2}) $ and $Q_{\text{out}} =  Q_{\text{out}}^{4 \rightarrow 1}  =-( E_{1} - E_{4}) = \alpha V_{1} (P_{4} - P_{1})$. The   relation for the two adiabats $1 \to 2$ and $3\to 4$  is
\begin{equation}
\label{eqn:otto1}
 P_{2}V_{2}^{\frac{\alpha + 1}{\alpha}}  =P_{1}V_{1}^{\frac{\alpha + 1}{\alpha}}\,,\qquad P_{3}V_{2}^{\frac{\alpha + 1}{\alpha}} = P_{4}V_{1}^{\frac{\alpha + 1}{\alpha}} \,.
\end{equation}
Dividing these two equations yields that the pressure ratios are equal: $P_3/P_2= P_4/P_1$. Using this equality  and  \eqref{eqn:otto1}    we find the   Otto efficiency 
\begin{equation}
    \label{eqn: otto_eff}
    \eta_{\text{Otto}} = 1- \frac{  Q_{\text{out}}^{4\rightarrow 1} }{ Q_{\text{in}
    }^{2\rightarrow 3} } = 1 - \frac{E_{4}-E_{1}}{E_{3}-E_{2}} = 1 - \left ( \frac{V_{2}}{V_{1}} \right )^{\frac{1}{\alpha}}\,.
\end{equation}

\noindent \textbf{Diesel engine.}
In the Diesel cycle paths $1\rightarrow 2$ and $3\rightarrow 4$ are adiabats,   path $2\rightarrow 3$ is an isobar (which implies $P_{2} = P_{3}$),  and path $4\rightarrow 1$ is an isochore (which implies $V_{4}=V_{1}$). There is heat gain along $2\rightarrow 3$ and heat loss along $4\rightarrow 1$. For the isobar we use that the heat exchange is equal to the change in enthalpy, and for the isochore the heat exchange is equal to the change in internal energy. Thus, $Q_{\text{in}} = Q_{\text{in}}^{2 \rightarrow 3} = H_{3} - H_{2} = (\alpha+1) P_{2}(V_{3}-V_{2}) $ and $Q_{\text{out}} =  Q_{\text{out}}^{4 \rightarrow 1} = -(E_{1} - E_{4}) = \alpha  V_{1}(P_{4}-P_{1}) $. Moreover, the  two adiabats $1\rightarrow 2$ and $3\rightarrow 4$ satisfy the  relations
\begin{equation}
\label{eqn:adia-diesel1}
     P_{1}V_{1}^{\frac{\alpha + 1}{\alpha}}\, =P_{2}V_{2}^{\frac{\alpha + 1}{\alpha}} , \qquad 
    P_{2}V_{3}^{\frac{\alpha + 1}{\alpha}} = P_{4}V_{1}^{\frac{\alpha + 1}{\alpha}}\,.
\end{equation}
Inserting these adiabat relations into the equations for the heat transfers   we obtain the Diesel efficiency 
\begin{equation}
    \label{eqn: diesel_eff}
    \eta_{\text{Diesel}} =1-\frac{Q_{\text{out}}^{4 \rightarrow 1}}{Q_{\text{in}}^{2 \rightarrow 3} }= 1 - \frac{E_{4}- E_{1}}{H_{3}- H_{2}} = 1 - \frac{\alpha}{\alpha + 1}\left ( \frac{V_{2}}{V_{1}} \right )^{\frac{1}{\alpha}}\frac{\left ( \frac{V_{3}}{V_{2}} \right )^{\frac{\alpha +1}{\alpha}}-1}{\left ( \frac{V_{3}}{V_{2}} \right )-1}\,.
\end{equation}
We checked     by plotting in various dimensions that the CFT efficiency is less than the efficiency for a monatomic and diatomic ideal gas. \\

\noindent \textbf{Rectangle engine.}
In this cycle paths $2\rightarrow 3$ and $4\rightarrow 1$ are isobars, i.e., $P_2=P_3$ and $P_4=P_1$, and paths $1\rightarrow 2$ and $3\rightarrow 4$   are isochores, i.e., $V_1=V_2$ and $V_3=V_4$. There is an inward flow of heat along $1\rightarrow 2$ and $2\rightarrow 3$, whereas there is an outward flow of heat along the paths $3\to 4$ and $4 \to 1$. Along the isobars the heat input is equal to the enthalpy difference, and along the isochores the heat change is equal to the internal energy difference. Thus, the total heat that flows into the system is given by
\begin{equation}
\label{eqn:ii}
    Q_{\text{in}} = Q_{\text{in}}^{1\rightarrow 2}+Q_{\text{in}}^{2\rightarrow 3}  =   (E_{2} - E_{1})+(H_{3} - H_{2})=\alpha V_1 (P_2 - P_1) + (\alpha + 1)P_2 (V_4 - V_1)\,.
\end{equation}
For this engine the work produced can be   easily calculated, since it is  the area enclosed by the cycle
\begin{equation}
    W = (P_{2}-P_{1})(V_{4}-V_{1})\,.
\end{equation}
Thus, substituting for enthalpy and internal energy in \eqref{eqn:ii}, we get the following efficiency
\begin{equation}
\label{eqn:e5}
    \eta_{\text{rectangular}} =\frac{W}{Q_{\text{in}}}= \frac{1}{(\alpha + 1)\left ( \frac{P_{2}}{P_{2}-P_{1}} \right ) + \alpha\left ( \frac{V_{1}}{V_{4}-V_{1}} \right )}\,.
\end{equation}
The CFT engine is less efficient than the corresponding ideal gas engine if $\gamma > \frac{D}{D-1} $.\\

\noindent \textbf{Stirling engine.} 
In a Stirling cycle paths $1\rightarrow 2$ and $3\rightarrow 4$ are isotherms and paths $2\rightarrow 3$ and $4\rightarrow 1$ are isochores (hence $V_2=V_3$ and $V_4=V_1$). In the absence of a regenerator, there  is heat gain along paths $1 \rightarrow 2$ and $ 4 \rightarrow 1$, and there is heat loss along the paths $2 \rightarrow 3 $ and $3 \rightarrow 4 $. The heat transfer along the isochores is equal to the change in internal energy, and the heat exchange along the isotherm follows from the Clausius relation \eqref{eqn:heat_}. Thus,  $Q_{\text{in}} = Q_{\text{in}}^{1\rightarrow 2} + Q_{\text{in}}^{4\rightarrow 1} = T_{\text{h}}(S_{2}-S_{1}) + \alpha V_{1}(P_{1}-P_{4}) $ and $Q_{\text{out}} = Q_{\text{out}}^{2\rightarrow 3} + Q_{\text{out}}^{3\rightarrow 4} =  - \alpha V_{2}(P_{3}-P_{2})-T_{\text{c}}(S_{4}-S_{3}) $. Hence, the Stirling efficiency can be written as follows
\begin{equation}
    \label{eqn: stirling_eff}
    \eta_{\text{Stirling}} = 1-\frac{Q_{\text{out}}^{2\to 3}+Q_{\text{out}}^{3\to 4}}{Q_{\text{in}}^{1\to 2}+ Q_{\text{in}}^{4\to 1}}=1 - \frac{T_{\text{c}}(S_{3}-S_{4})+ \alpha V_{2}(P_{2}-P_{3})}{T_{\text{h}}(S_{2}-S_{1})+ \alpha V_{1}(P_{1}-P_{4})}\,.
\end{equation}

\section{Equations for thermodynamic processes  in the $PV$- and $TS$-cycles}
\label{appb}

In this appendix  we explain how the $PV$- and $TS$-diagrams in Figures \ref{fig:five_figures} to \ref{fig:righT_column} are obtained for the CFT dual to the AdS-Schwarzschild black hole and the ideal gas. To construct the $PV$ cycles for various engines, it is necessary to determine the equations of the various paths, namely the adiabat, isotherm, isochore, and isobar. In the $P-V$ plane, the equations for an  isobar and isochore are simply   $P = \text{const}.$ and $V = \text{const}.$, respectively. Further, the equation for an adiabat in a CFT is  given by
\begin{equation}
    \label{eqn: CFT_PV_eqn_adiabat}
    PV^{\frac{D}{D-1}} = \text{constant} \qquad \text{(adiabat for CFT)}\,.
\end{equation} 
This adiabat relation can be derived from the first law \eqref{quasistatic} and the equation of state \eqref{eqn:eqn_of_state}. Both  volume and pressure vary along the adiabat, hence the equation of state yields $\Delta E = \alpha(P\Delta V + V \Delta P)$. Since there is no heat exchange along an adiabat ($Q = 0$), it follows from the   first law that, $(\alpha + 1)P \Delta V + \alpha V \Delta P = 0.$ Dividing both sides by $\alpha PV $ and integrating we arrive at \eqref{eqn: adiabat_relation}. Finally, inserting $\alpha =D-1$ for a CFT into \eqref{eqn: adiabat_relation} yields \eqref{eqn: CFT_PV_eqn_adiabat}.

To determine the equation of the isotherm for the holographic CFT engine, we first  solve for $x$ in terms of $T$ and $V$ from \eqref{eqn:holo_temp1}, see equation \eqref{eqn:x_T_R}, and substitute $x(T,V)$ in the expression for pressure  $P(V,C,x)$, see equation \eqref{holo:pressure}. Further, since temperature is constant along an isotherm, we fix the temperature to $T_o$ (and we also fix the central charge  $C$). We arrive at the following equation for the isotherm
\begin{align}
\label{eqn: CFT_PV_isotherm}
    &PV^{\frac{D}{D-1}} = C (\Omega_{D-1})^{\frac{1}{D-1}} 
    \left[\frac{1}{D} \left(2 \pi T_o \left(\frac{V}{\Omega_{D-1}}\right)^{\frac{1}{D-1}} 
    + \sqrt{ 4 \pi ^{2} T_o^2\left(\frac{V}{\Omega_{D-1}}\right)^{\frac{2}{D-1}}- D(D-2)}\right)\right]^{D-2}\times \\
    &\quad \times 
    \left[
    1 + 
     \left(\frac{1}{D} \left(2 \pi T_o  \left(\frac{V}{\Omega_{D-1}}\right)^{\frac{1}{D-1}} + \sqrt{  4 \pi ^{2} T_o^2 \left(\frac{V}{\Omega_{D-1}}\right)^{ \frac{2}{D-1}}- D(D-2)}\right)\right)^{2} 
    \right]\qquad \text{(isotherm for holographic CFT)}\,.\nonumber
\end{align}
With the equations for all the thermodynamic processes at hand, we can obtain the $PV$-plots for all the holographic heat engines in Appendix \ref{appa}.

For an ideal gas  the equation for an  adiabat is  
\begin{equation}
    \label{eqn: IG_adiabat}
    PV^{\gamma} = \text{constant} \qquad \text{(adiabat for ideal gas)}\,.
\end{equation}
Moreover, the equation of the isotherm for an ideal gas is given by its equation of state
\begin{equation}
    \label{eqn:IG_isotherm}
    PV = NT_{o} \qquad \text{(isotherm for ideal gas)}\,.
\end{equation}
We also derive an equation for $\Delta S$ for an ideal gas along an isotherm, since we use that in the main text in computing   the Stirling efficiency. We first note that the change in the internal energy of an ideal gas along an isotherm is zero, due to   equipartition     $E= \frac{f}{2} N T$ (if we fix $N$). Hence,  the first law implies $ Q = P \Delta V$  along an isotherm. Substituting the Clausius relation $ Q = T \Delta S$ and the equation of state $P = NT/V$ in the first law and integrating on both sides, yields $\Delta S = N \log(V_2/V_1). $

Next, we determine the equations for various paths in the $TS$-diagrams. In the $T-S$ plane, the equations for   adiabats and isotherms are simply   $T = \text{const}.$ and $S = \text{const}.$, respectively. To determine the equation for the isochore for thermal CFT systems dual to an AdS-Schwarzschild black hole, we make use of expression \eqref{eqn:holo_temp1} for the temperature and substitute $x$ in terms of $S$ from \eqref{eqn:central_charge} and further fix the volume to $V_o$, yielding
\begin{equation}
    \label{TS_CFT_Isochore}
     T = \frac{D-2}{4 \pi}\left(\frac{\Omega_{D-1}}{V_o}\right)^{\frac{1}{D-1}}\left(\frac{4 \pi C}{S}\right)^{\frac{1}{D-1}}\left (1 + \frac{D}{D-2}\left(\frac{S}{4 \pi C}\right)^{\frac{2}{D-1}} \right)
\qquad \text{(isochore for holographic CFT)}\,.\end{equation}
To obtain the equation for an  isobar, we first find $V$ in terms of $P$,  $S$  and $C$ by combining the expressions \eqref{holo:pressure} and~\eqref{eqn:central_charge} for the pressure $P(V,C,x)$  and $x(S,C)$, respectively,  and solving for $V$. Then we substitute $V(P,S,C)$   into \eqref{TS_CFT_Isochore} and fix the pressure to $P_o$. Finally, the isobar equation in the $T-S$ plane becomes
\begin{equation}
    \label{eqn:TS_CFT_Isobar}
    T = \frac{D-2}{4 \pi} \left(\frac{P_o \Omega_{D-1}}{C}\right)^{\frac{1}{D}} \frac{\left[ 1 + \frac{D}{D-2} \left(\frac{S}{4 \pi C}\right)^{\frac{2}{D-1}} \right]}{\left[\left(\frac{S}{4 \pi C}\right)^{2} \left(1 + \left(\frac{S}{4 \pi C}\right)^{\frac{2}{D-1}}\right)\right]^{\frac{1}{D}}}\qquad \text{(isobar for holographic CFT)}\,.
\end{equation}
The  isochore equation for the ideal gas case can be obtained from equation (1a) on page 19 of \cite{Pathria:1996hda} 
\begin{equation}
    \label{eqn:TS_IG_Isochore}
    T = \exp\left[\frac{(\gamma - 1)S}{ N} - \gamma \right] \left(\frac{V_o}{N}\right)^{-(\gamma - 1)}\frac{1}{2\pi m} \qquad \text{(isochore for ideal gas)}\,.
\end{equation}
where $m$ is the mass of the gas particles. The  isobar equation can be   obtained by writing $V_0$ in terms of $P_0$ in \eqref{eqn:TS_IG_Isochore}, using the ideal gas equation of state  $V/N= T/P$,
\begin{equation}
    \label{eqn: TS_CFT_Isobar}
    T^{\gamma} = \exp\left[\frac{(\gamma -1)S}{N} - \gamma \right] P_o^{\gamma - 1} \frac{1}{2 \pi m} \qquad \text{(isobar for ideal gas)}\,.
\end{equation}
These expressions allow us to  obtain the $TS$-diagrams for various holographic and ideal gas heat engines.  

\section{High-temperature expansion of the  Stirling efficiency  for a     CFT   on a sphere}
\label{appc}

Consider a general conformal field theory on a round sphere $    S^{D-2}$ at finite temperature.  The perturbative expansion of the canonical free energy $F$ around $TR = \infty$ takes the form  \cite{Kutasov:2000td}
\begin{equation}
    - F R = a_D (2\pi TR)^D + a_{D-2} (2\pi TR)^{D-2}+ a_{D-4}(2\pi T R )^{D-4} + \cdots  \,.
\end{equation} 
At strong 't Hooft coupling the high-temperature expansion
of the free energy includes an infinite series in $1/(TR)$, whereas at zero coupling the coefficients of all the terms that scale with negative powers of $TR$ are absent. Non-perturbative corrections appear at order $\mathcal O(e^{-(2
\pi)^2 TR})$, which we ignore. Below we compute the Stirling efficiency in the high-temperature or large-volume expansion up to subsubleading order, i.e., keeping the coefficients $a_D, a_{D-2}$ and $a_{D-4} $ finite and neglecting higher-order corrections.

The entropy $S$, energy $E$ and pressure $P$ can be derived from the free energy as follows
\begin{align}
\label{entropyexpan}
    S &= - \left (  \frac{\partial F}{\partial T} \right)_V = a_D D 2\pi (2\pi  TR)^{D-1} + a_{D-2} (D-2) 2\pi (2\pi  TR)^{D-3} + a_{D-4}(D-4)2\pi(2\pi TR)^{D-5}+ \cdots \\
    ER&=FR - TR \left (  \frac{\partial F}{\partial T} \right)_V= a_D (D-1) (2\pi  TR)^{D} + a_{D-2} (D-3) (2\pi  TR)^{D-2}+ a_{D-4}(D-5)   (2\pi TR)^{D-4}+\cdots \label{energyexpan}\\
    P&=-\left ( \frac{\partial F}{\partial V} \right)_T= \frac{1}{(D-1)V R} \left ( a_D (D-1)(2\pi T R)^D    + a_{D-2}(D-3) (2\pi   TR)^{D-2}  +a_{D-4} (D-5) (2\pi T R )^{D-4} + \cdots \right)\,.
\end{align}
Inserting these functions $S(T,V)$ and $P(T,V)$ into the Stirling efficiency \eqref{eqn: stirling_eff1a} yields
\begin{equation}  \eta_{\text{Stirling}}^{\text{CFT}} = 1 - \frac{T_{\text{c}}^{D}(V_{2}\xi_{12}-V_{1}\xi_{11})+V_{2}\frac{D-1}{D}(T_{\text{h}}^{D}\chi_{22}-T_{\text{c}}^{D}\chi_{12})}{T_{\text{h}}^{D}(V_{2}\xi_{22}-V_{1}\xi_{21})+V_{1}\frac{D-1}{D}(T_{\text{h}}^{D}\chi_{21}-T_{\text{c}}^{D}\chi_{11})},
\end{equation}
where $\xi_{ij}$ and $\chi_{ij}$ are up to order $\mathcal O (T_i^{-6} V_j^{-6/(D-1)})$
\begin{align}
    \xi_{ij} &= 1 + \frac{a_{D-2}(D-2)}{a_D D (2\pi)^2 T_i^2} \left ( \frac{\Omega_{D-1}}{V_j} \right)^{\frac{2}{D-1}}+ \frac{a_{D-4}(D-4)}{a_D D (2\pi)^4 T_i^4 }  \left ( \frac{\Omega_{D-1}}{V_j} \right)^{\frac{4}{D-1}} + \cdots \label{xiappendix}\\
    \chi_{ij} &= 1 + \frac{a_{D-2}(D-3)}{a_D (D-1)(2\pi)^2T_i^2}\left ( \frac{\Omega_{D-1}}{V_j} \right)^{\frac{2}{D-1}} + \frac{a_{D-4}(D-5)}{a_D (D-1)(2\pi)^4T_i^4}\left ( \frac{\Omega_{D-1}}{V_j} \right)^{\frac{4}{D-1}} + \cdots \label{chiappendix}
\end{align}
Here $T_{1} \equiv T_{\text{c}}$ and $T_{2} \equiv T_{\text{h}}$. 
The subleading corrections to the Stirling efficiency thus depend on the coefficients in the free energy expansion, which are fixed by the matter content of the CFT. For free CFTs in $D=4$ with $n_S$ scalars, $n_F$ Weyl fermions and $n_V$  vector fields the free energy coefficients are given by \cite{Kutasov:2000td} 
\begin{align} 
    a_4 &= \frac{1}{720} \left ( n_S + 2 n_V + \frac{7}{4} n_F \right)\,,\\
    a_2 &= -\frac{1}{24} \left ( 2 n_V + \frac{1}{4} n_F \right)\,,\\
    a_0 &= \frac{1}{240}\left ( n_S + 22 n_V + \frac{17}{4} n_F\right)\,.
\end{align}
For instance,  $\mathcal N=4$ SYM theory with $\mathrm{SU(}N\mathrm{)}$  gauge group has $6n$ scalars, $4n$ Weyl fermions and $n$ gauge bosons, with $n=N^2-1$, so these coefficients are in this case: $a_4 = n/48, a_2=-n/8$ and $a_0=3n/16$, hence $a_2 = - 6 a_4$ and $a_0 = 9a_4$.
Moreover, for $D$-dimensional holographic CFT states dual to a $(D+1)$-dimensional AdS-Schwarzschild black hole the coefficients are  
\begin{equation} \label{holographiccoeff1}
    \frac{a_{D-2}}{a_D} = - \frac{D^2 (D-1)}{4}\qquad \text{and}\qquad \frac{a_{D-4}}{a_D} =  \frac{D^3 (D-2)^2(D-1)}{32}\,,
\end{equation}
since with this choice of coefficients the expressions \eqref{xiappendix}-\eqref{chiappendix} match with \eqref{eqn: xi_prime}-\eqref{eqn:chi_prime} below. For instance, in $D=4$ we have $a_2=-12 a_4$, $a_0= 24a_4$. Assuming these formulae are valid at strong coupling in $\mathcal N=4$ SYM in $D=4$, the coefficients flow modestly from weak to strong coupling. It is well known that the leading coefficient at strong 't Hooft coupling $\lambda(\equiv g^2N)=\infty$ is $3/4$ times the leading coefficient at weak coupling $\lambda=0$ \cite{Gubser:1996de}. This implies for the coupling dependence of the other coefficients:
\begin{equation}
    a_4 (\lambda=\infty) =\frac{3}{4} a_4(\lambda=0)\,,\qquad a_2 (\lambda=\infty) =\frac{3}{2} a_2(\lambda=0)\,,\qquad a_0 (\lambda =\infty) =2a_0(\lambda=0)\,.
\end{equation}
The second relation was also derived in \cite{Kutasov:2000td}. 

Finally, in order to relate    the high-temperature expansion in \cite{Kutasov:2000td} with the large entropy expansion in \cite{Verlinde:2000wg}, we express $TR$ and $ER$ as a function of   $S$. We  first solve \eqref{entropyexpan} for $TR$ and then insert  this expression into \eqref{energyexpan}, yielding
\begin{align}
TR&= \frac{1}{(a_D D (2\pi)^D)^{\frac{1}{D-1}}}S^{\frac{1}{D-1}}  - \frac{ (D-2)a_{D-2}}{(D-1)(a_D2\pi D)^{\frac{D-2}{D-1}}} S^{-\frac{1}{D-1}}+\cdots\,, \\
ER &=\frac{D-1}{ (a_D(2\pi D)^D)^{\frac{1}{D-1}}}S^{\frac{D}{D-1}} -\frac{ a_{D-2}}{(a_D2\pi D)^{\frac{D-2}{D-1}}} S^{\frac{D-2}{D-1}}+ \cdots\,.
\end{align}
Comparing this to the large entropy expansion in \cite{Verlinde:2000wg} \begin{equation}ER = \frac{a}{4\pi} S^{\frac{D}{D-1}} + \frac{b}{4\pi}S^{\frac{D-2}{D-1}}\,,\end{equation} we can read off the relation between $(a_D,a_{D-2})$ and $(a,b)$. In particular, the ratio of $a_{D-2}$ and $a_D$ is proportional to the product $ab$  \begin{equation}\frac{a_{D-2}}{a_D} = - \frac{ab D^2}{4(D-1)}\,.\end{equation} Hence, the holographic result \eqref{holographiccoeff1} for the free energy coefficients translates into $ab=(D-1)^2$, consistent with \cite{Verlinde:2000wg}.

\section{Holographic dictionary for the thermodynamics of AdS-Schwarzschild black holes}
\label{appd}

In this appendix we derive the holographic dictionary for the thermodynamic variables of AdS-Schwarzschild black holes, which are solutions to the Einstein equation with a negative cosmological constant. The line element of   AdS-Schwarzschild geometry  in static coordinates in $D+1$ dimensions is
\begin{equation} \label{adsschw}
    ds^2= - f(r)dt^2 + \frac{dr^2}{f(r)}+r^2d\Omega_{D-2}^2\,,
\end{equation}
with blackening factor
\begin{equation}
    f(r)= 1+\frac{r^2}{L^2}- \frac{m}{r^{D-2}}\,, \qquad \text{where} \qquad m = r_h^{D-2} \left ( 1 + \frac{r_h^2}{L^2}\right)
\end{equation}
is the mass parameter and $r_h$ is the horizon radius. The AdS curvature radius $L $ is related to the cosmological constant by: $\Lambda = - D (D-1)/ (2L^2)$. The mass $M$, Bekenstein-Hawking entropy $S$ and Hawking temperature  $T_{\text{H}}$ for an AdS-Schwarzschild black hole are \cite{Bekenstein:1973ur,Bardeen:1973gs,Hawking:1975vcx,Hawking:1982dh} 
\begin{align}
    M&= \frac{(D-1)\Omega_{D-1}m}{16 \pi G}=\frac{(D-1)\Omega_{D-1}}{16 \pi G}r_h^{D-2} \left ( 1 + \frac{r_h^2}{L^2}\right)\,,\label{massblackhole}\\
     S &= \frac{A(r_h)}{4G} = \frac{\Omega_{D-1} r_h^{D-1}
}{4G}\,,\label{Bekensteinhawking}\\
T_{\text{H}} &=   \frac{|f'(r_h)|}{4\pi}= \frac{Dr_h^2 + (D-2)L^2}{4\pi r_h L^2 } \,. \label{hawkingtemperature}
\end{align}
The mass is defined through background subtraction so that pure AdS spacetime has $M=0.$
These thermodynamic variables satisfy the first law of black hole mechanics: $dM = T_{\text{H}} dS.$ The thermodynamics of an  AdS-Schwarzschild black hole is dual to the thermodynamics of a holographic CFT. The CFT lives on the conformal boundary of AdS-Schwarzschild geometry, i.e., the CFT metric is a Weyl rescaling of the asymptotic geometry $g_{\text{CFT}} = \lim_{r\to \infty} \Omega^2(x) g_{\text{AdS}}$ \cite{Gubser:1998bc,Witten:1998qj}. To leading order in an expansion around $r 
= \infty $ the line element \eqref{adsschw} becomes
\begin{equation}
    ds^2 = -\frac{r^2}{L^2} dt^2  + \frac{L^2}{r^2} dr^2 + r^2 
    d\Omega_{D-1}^2.
\end{equation}
For the Weyl factor we choose $\Omega = R/r$, so that the CFT line element reads
\begin{equation}
    ds^2_{\text{CFT}} = - \frac{R^2}{L^2} dt^2 + R^2 d \Omega_{D-1}^2\,.
\end{equation}
The boundary spatial volume is  thus $V = \Omega_{D-1} R^{D-1}$. Moreover,  the time variable in this Weyl frame is $R/L$ times the global AdS time $t.$ This factor also appears in the dictionary for the CFT energy $E$ and temperature $T$. They are, respectively, identified with the mass and Hawking temperature of the black hole times the inverse of this factor  \cite{Savonije:2001nd,Visser:2021eqk,Ahmed:2023snm}
\begin{equation} \label{dictionaryeandt}
    E = M\frac{L}{R} \,,\qquad T = T_{\text{H}}\frac{L}{R}\,.
\end{equation}
Now, the CFT thermodynamic variables satisfy a thermodynamic first law that is dual to the first law of AdS-Schwarzschild black holes,
\begin{equation}
    \Delta E = T \Delta S - P\Delta V\,.
\end{equation}
Crucially, for the boundary and bulk first laws to match,  the pressure $P$ must satisfy  the conformal equation of state
\begin{equation}\label{eqnstatepressure}
P=\frac{E}{(D-1)V}\,.
\end{equation}
Further, the thermodynamic entropy in the CFT is identified with the Bekenstein-Hawking entropy~\eqref{Bekensteinhawking}. Finally, we introduce the holographic dictionary for the central charge $C$ (that holds for Einstein gravity with a negative cosmological constant) and the dimensionless parameter $x$ \cite{Henningson:1998gx,Myers:2010xs}
\begin{equation}\label{centralchargedic}
    C=\frac{\Omega_{D-1}L^{D-1}}{16 \pi G}\,, \qquad x = \frac{r_h}{L}\,.
\end{equation}
This central charge is defined in the CFT as the dimensionless proportionality factor of the energy and entropy in the canonical ensemble, i.e., $E=C f_E(T,V)$ and $S=C f_S(T,V)$ \cite{Witten:1998zw,Klemm:2001db,Visser:2021eqk}.  The CFT thermodynamic variables  can be expressed in terms of   $C,x$ and $R $ (or $V$) by combining the equations \eqref{massblackhole}, \eqref{Bekensteinhawking}, \eqref{hawkingtemperature}, \eqref{dictionaryeandt}, \eqref{eqnstatepressure} and \eqref{centralchargedic} in this appendix. The resulting holographic dictionary for $S,E,T$, and $P$ is given in  \eqref{eqn:central_charge}-\eqref{holo:pressure}.

\section{Holographic Stirling efficiency as a function of temperature and volume}
\label{appe}

In this appendix we obtain the   efficiency of the holographic Stirling engine as a function of temperature $T$ and volume $V$. To achieve this, we   express the entropy $S$ and pressure $P$ in terms of $T$ and $V$ for thermal CFT states dual to AdS-Schwarzschild black holes. First,  we solve \eqref{eqn:holo_temp1} for  the parameter $x$ (which is related to $S$) in terms of $T$ and $R$ 
\begin{equation}
    \label{eqn:x_T_R}
    x = \frac{1}{D} \left(2 \pi TR + \sqrt{ 4 \pi ^{2}T^{2}R^{2}- D(D-2)  }\right)\,.
\end{equation} 
An important detail is that we take the plus sign in front of the square root, since this corresponds to   large AdS black holes with $x > \sqrt{(D-2)/D}$, which have positive heat capacity $c\equiv T (\partial S/ \partial T)_{V,C}$ \cite{Hawking:1982dh} (small black holes with $x < \sqrt{(D-2)/D}$ have  negative heat capacity). We need solutions with positive heat capacity for the thermodynamic cycles in the $PV$-plots to act as heat engines. 
We obtain $S(T,V)$ by   substituting \eqref{eqn:x_T_R} in \eqref{eqn:central_charge} and   using the relation between $R$ and the volume, $V=\Omega_{D-1}R^{D-1}$. The resulting expression is   
\begin{equation}
    \label{eqn:S_holo_T_V_exact}
    S = 4 \pi C \left[  \frac{1}{D} \left(2 \pi T \left(\frac{V}{\Omega_{D-1}}\right)^{\frac{1}{D-1}} + \sqrt{ 4 \pi ^{2}T^{2} \left(\frac{V}{\Omega_{D-1}}\right)^\frac{2}{D-1}- D(D-2)}\right)   \right]^{D-1}\,.
\end{equation}
Further, $P(T,V)$ is   obtained by inserting \eqref{eqn:x_T_R} into the dictionary \eqref{holo:pressure} for pressure.   The result  is given in equation~\eqref{eqn: CFT_PV_isotherm}, with $T_{o}$ replaced by the general temperature $T$. 
Substituting $S(T,V)$ and $P(T,V)$ in the   expression \eqref{eqn: stirling_eff1a} for the Stirling efficiency, we obtain the exact efficiency of the holographic Stirling engine in terms of temperature and volume 
\begin{equation}
   \label{eqn:stirling_holo_exCt_eff}
    \eta^{\text{black hole}}_{\text{Stirling}} = 1 - \frac{T_{c}^{D}(V_2 \xi_{12} - V_1 \xi_{11}) + V_2 \frac{D-1}{ D}(T_{\text{h}}^{D} \chi_{22} - T_{c}^{D}\chi_{12})}{T_{\text{h}}^{D}(V_2 \xi_{22} - V_1 \xi_{21}) + V_1 \frac{D-1}{ D}(T_{\text{h}} ^{D} \chi_{21} - T_{\text{c}}^{D} \chi_{11})} \,,
\end{equation}
where $\xi_{ij}$ and $\chi_{ij}$ are defined as
\begin{equation}
    \label{eqn:kappa}
    \xi_{ij} = \frac{1}{2^{D-1}}\left[ 1 + \sqrt{1 - \frac{D(D-2)}{4 \pi^{2} T_{i}^{2}} \left(\frac{\Omega_{D-1}}{V_j}\right)^{\frac{2}{D-1}}}  \right]^{D-1} \,\!\!\!\!\!\!\!\!\!\!, \qquad \chi_{ij} = \xi_{ij} ^{\frac{D-2}{D-1}} \left[ \xi_{ij}^{\frac{2}{D-1}} + \frac{D^{2}}{16 \pi^{2} T_i^{2}} \left(\frac{\Omega_{D-1}}{V_j}\right)^{\frac{2}{D-1}} \right]\,.
\end{equation}
Here $T_{1} \equiv T_{\text{c}}$ and $T_{2} \equiv T_{\text{h}}$.  With these expressions in hand, we can also compute subleading corrections to the Stirling efficiency in the high-temperature or large-volume expansion. In order to achieve this we first expand $S(T,V)$ and $P(T,V)$ up to order $\mathcal O(T_i^{-6} V_j^{-6/(D-1)})$
\begin{align}
\label{eqn:entro_second_order_exp}
    &S = \frac{4 \pi C}{\Omega_{D-1}} \left( \frac{4 \pi}{D}\right)^{D-1}T^{D-1}V \left [ 1- \frac{D(D-1)(D-2)}{16\pi^{2}T^{2}} \left(\frac{\Omega_{D-1}}{V}\right)^{\frac{2}{D-1}} + \frac{D^{2}(D-1)(D-2)^{2}(D-4)}{512 \pi^{4} T^{4}} \left(\frac{\Omega_{D-1}}{V}\right)^{\frac{4}{D-1}} + \cdots\right] \\
    \label{eqn:pressure_second_order_exp}
    &P = \frac{C}{\Omega_{D-1}} \left(\frac{4 \pi}{D}\right)^{D} T^{D} \left[1 - \frac{D^{2}(D-3)}{16 \pi^{2} T^{2}}\left(\frac{\Omega_{D-1}}{V}\right)^{\frac{2}{D-1}} + \frac{D^{3}(D-2)^{2}(D-5)}{512 \pi^{4}T^{4}} \left(\frac{\Omega_{D-1}}{V}\right)^{\frac{4}{D-1}} + \cdots \right]\,.
\end{align}
Inserting  these two expressions in \eqref{eqn: stirling_eff1a}, we find the subleading corrections to 
  $\xi_{ij}$ and $\chi_{ij}$   up to order $\mathcal O(T_i^{-6} V_j^{-6/(D-1)})$
\begin{align}
    \label{eqn: xi_prime}
    &\xi_{ij} = 1- \frac{D(D-1)(D-2)}{16\pi^{2}T_i^{2}} \left(\frac{\Omega_{D-1}}{V_j}\right)^{\frac{2}{D-1}} + \frac{D^{2}(D-1)(D-2)^{2}(D-4)}{512 \pi^{4} T_i^{4}} \left(\frac{\Omega_{D-1}}{V_j}\right)^{\frac{4}{D-1}} +\cdots\,,\\
    \label{eqn:chi_prime}
    &\chi_{ij} = 1 - \frac{D^{2}(D-3)}{16 \pi^{2} T_i^{2}}\left(\frac{\Omega_{D-1}}{V_j}\right)^{\frac{2}{D-1}} + \frac{D^{3}(D-2)^{2}(D-5)}{512 \pi^{4}T_i^{4}} \left(\frac{\Omega_{D-1}}{V_j}\right)^{\frac{4}{D-1}}+\cdots\,.
\end{align}
This agrees with expanding \eqref{eqn:kappa} up to subsubleading order around $TR =\infty$. Comparing this to the (sub)subleading corrections to the Stirling efficiency for a general CFT in \eqref{xiappendix}-\eqref{chiappendix}, we see that the free energy coefficients  are given by \eqref{holographiccoeff1} for holographic CFTs.

\end{document}